\documentclass[preprint2]{aastex}
\usepackage{graphicx}
\usepackage{amssymb}
\usepackage{amsmath}
\newcommand{\psr}{PSR B0656+14}

\begin{document}
\title{Multiwavelength spectroscopy of \psr}

\author{Martin Durant$^1$, Oleg Kargaltsev$^1$ and George G. Pavlov$^{2,3}$}
\affil{$^1$ University of Florida, 211 Bryant Space Science Center, Gainesville, FL, USA\\
$^2$ Pennsylvania State University, 525 Davey Lab, University Park, PA, USA\\
$^3$ St.-Petersburg State Polytechnic University, Polytekhnicheskaya ul.\ 29,
195251, Russia}
\email{martin.durant@astro.ufl.edu}

\begin{abstract}
Using high-quality Hubble Space Telescope observations, we construct the near infra-red (NIR) to far ultra-violet (FUV) spectral energy distribution (SED) of \psr. The SED is non-monotonic. Fitting it with a simple combination of a Rayleigh-Jeans spectrum (UV) and non-thermal power-law (optical/NIR) leaves significant residuals, strongly hinting at one or more spectral features. We consider various models (combination of continuum components, and absorption/emission lines) with possible interpretations, and place them in the context of the broader spectral energy distribution. Surprisingly, the extrapolation of the best-fit X-ray spectral model  roughly match the NIR-FUV data, and the power-law component is also consistent with the $\gamma$-ray fluxes. We compare the multiwavelength SED  of B0656+14 with those of other optical, X-ray and $\gamma$-ray detected pulsars, and notice that a simple power-law spectrum crudely accounts for most of the non-thermal emission.
\end{abstract}
\maketitle

\section{Introduction}
Pulsars are responsible for the production of high-energy emission observable across the EM spectrum from the radio to  $\gamma$-rays. They are natural laboratories for investigating particle acceleration and radiation processes in the high-energy, high-field, high-gravity regime. 
Outside the radio, the non-thermal high energy emission is believed to be produced by incoherent processes such as synchrotron and curvature radiation. With current instrumentation, this radiation is most easily detected in $\gamma$-rays (where most of the energy is released) and X-rays (although the latter is often ``contaminated'' by thermal emission from the neutron star surface). It is much more challenging to detect the non-thermal emission in the optical/near-infrared (NIR).

There are few pulsars for which
multicolor UV/optical/IR photometry exists \citep{2009arXiv0908.1010M,2007Ap&SS.308..287K,2007A&A...473..891M}. These are moderately old pulsars B1929+10 and B0950+08; $\tau\simeq3$\,Myr and 17.5\,Myr, respectively),
 middle-aged pulsars Geminga, B0656$+$14, and B1055$-$52  ($\simeq 300$, 100, and 500\,kyr respectively), the younger Vela pulsar
($\simeq 10$ kyr), and the very young pulsars Crab (950 yrs)
and B0540$-$69 ($\simeq 1.7$ kyr). The optical and X-ray spectra of
the Vela, Crab and B0540$-$69, look ``boring'' (featureless power-laws), while
the spectra of the  middle-aged pulsars 
are more complex and exhibit both thermal (emitted from the NS
surface) and nonthermal (magnetospheric) components in X-rays and optical-UV (e.g., \citealt{2005ApJ...627..383R}). No  spectral feature has yet been observed in the IR-UV spectra of pulsars\footnote{\citet{1996ApJ...456L.111B} claimed a feature in the optical-UV spectrum of Geminga (see also \citealt{1998A&A...332L..37M}), but this was not supported by subsequent observations \citep{2005ApJ...625..307K}}, although \citet{2007Ap&SS.308..545Z}'s data hinted at possible feature(s) for \psr.
Here we will present detailed analysis of the IR-UV spectrum of B0656+14.

PSR B0656+14 ($P=385$ ms; $\tau = 1.1\times 10^5$\,yr), located at
the distance of $288\pm30$ pc \citep{2003ApJ...593L..89B}, is the brightest of middle-aged pulsars, in
both X-rays and optical. Its X-ray spectrum and pulse profile were
studied by \citet{1993ApJ...414..867A,1996A&A...313..565P,1996ApJ...465L..35G,2002ApJ...574..377M,2004MmSAI..75..458Z,2005ApJ...623.1051D}. The X-ray spectrum (0.1--6 keV),
obtained from {\sl Chandra} observations \citep{2002nsps.conf..273P},
can be fitted with a model which consists of thermal soft (TS) component with a blackbody (BB) temperature of
$\simeq0.82$ MK, emitted from a large
part of the NS surface ($R\simeq7.3$ km); thermal hard (TH)
component with the BB temperature of $\simeq1.7$ MK, apparently emitted
from smaller ($R\simeq0.5$ km) hotter areas (perhaps hot polar
caps); and non-thermal power-law (PL) component, with the photon
index $\Gamma\simeq1.5$, possibly emitted from the pulsar magnetosphere.
 The X-ray radiation is pulsed, with a pulsed
fraction of $\sim13$\% at lower energies ($<0.7$ keV), where the TS
component dominates, and $\sim57\%$ in the $2-4$ keV band, where the
PL component is dominant.

The optical counterpart of \psr\ ($V\approx 25$)  was discovered by \citet{1994ApJ...422L..87C}. Because a nearby galaxy contaminates the ground-based images \citep{2001A&A...370.1004K}, the best near-IR/optical/near-UV
($\lambda = 0.2$--2 $\mu$m) data were obtained  with the {\sl Hubble
Space Telescope} ({\sl HST}). The pulsar was imaged with FOC \citep{1996ApJ...467..370P,1997ApJ...489L..75P},
WFPC2 \citep{2000ApJ...543..318M}, and NICMOS \citep{2001A&A...370.1004K}. These observations suggested that the pulsar has a
non-thermal spectrum from NIR through the optical, with somewhat
uncertain spectral index, $\alpha_{\nu}\simeq-0.4\pm0.5$
($F_{\nu}\propto\nu^{\alpha_{\nu}}$). The large uncertainty in the
slope reflects a substantial scatter among the existing photometric
points. This scatter could possibly be attributed to the presence of broad
spectral features or could be just
the result of poor cross-calibration between different instruments
that have been used to cover a broad range of wavelengths.

In addition to photometric observations, \citet{2003ApJ...597.1049K}
observed the optical pulse profile and polarization with the Palomar 5\,m telescope
 and found that the emission in 400$-$600\,nm band is
highly pulsed (upper limit on the un-pulsed flux is 16\%). The pulse
profile is double-peaked, with a bridge of emission between the two
peaks; emission from the bridge was found to be highly polarized.
Finally, the low-resolution spectrum and pulsations in the near-UV
(NUV) have been  measured with {\sl HST} STIS \citep{2005A&A...440..693S}. They found that NUV pulse profile is also double-peaked
and similar to the optical one. The high pulsed fraction, $70\pm12$
\%, and a flat spectrum ($\alpha_{\nu,{\rm NUV}}=0.35\pm0.5$) suggest that NUV emission is
 mostly non-thermal. Since the soft X-ray spectrum is thermal
 and optical spectrum is non-thermal, a transition between the thermal and
non-thermal spectra is expected to occur somewhere in between. \citet{2007Ap&SS.308..287K} show that the FUV flux is consistent with a low pulsed fraction thermal component and a highly pulsed non-thermal component. 

To better constrain the shape of the spectrum and
evaluate contributions of the thermal and non-thermal components, we
observed B0656 using instruments aboard {\sl HST} (ACS, STIS and COS). Together with re-analysis of archival NICMOS  observations, we present the most complete infrared to ultraviolet spectral energy distribution measured for any pulsar. The data analysis of the  observations, including the archival ones, is described in Section 2 and Appendix A. In Section 3, we present the results, both the photometry and spectra from the individual observations, and the combined spectral energy distribution. 
In Section 4 we discuss the interpretation of the measured spectral shape, its context in the wider spectral energy distribution, and comparisons with the spectra of other pulsars.

\section{Observations and analysis}
In Table \ref{log} we list the observations analyzed as part of this work. Of these, the NICMOS and STIS NUV/FUV observations were reported before \citep{2001A&A...370.1004K,2005A&A...440..693S,2007Ap&SS.308..287K}, but we reanalyze the data from the archive with the latest calibration applied. The rest of the data appear here for the first time.

Table \ref{old_phot}  summarizes previously published optical/UV observations of B0656+14, which we will not re-analyze in this paper. We will include the results of these observations with those produced in this work, for comparison.

\begin{deluxetable}{ccccc}
\tablecaption{Log of HST observations for analysis \label{log}}
\tabletypesize{\footnotesize}
\tablewidth{0in}
\tablehead{
\colhead{Date} & \colhead{Program ID} & \colhead{Instrument/} & \colhead{Wavelength}  & \colhead{Exposure} \\
& & \colhead{optical element} & \colhead{(\AA)} & \colhead{ time (s)}
}
\startdata 
1998-03-18 & 7836 & NICMOS/F110W & 11285 & 2544\\
 & 7836 & NICMOS/F160W &16056&  5089\\
 & 7836 & NICMOS/F187W & 18718 & 7633\\
2001-09-01 & 9156 &STIS/prism & 1750--3000 & 6791 \\
2001-11-16 & 9156 &STIS/prism & 1750--3000 & 12761\\
2004-01-20 & 9797 & STIS/G140L & 1150--1700 & 4950\\
2005-12-03 & 10600 & ACS/FR647M & 6820& 810\\
 & 10600 & ACS/FR647M & 7393& 810\\
 & 10600 & ACS/FR914M & 7956& 747\\
 & 10600 & ACS/FR914M & 8751& 1331\\
 & 10600 & ACS/FR914M & 9491& 2842\\
2005-12-08 & 10600 & ACS/FR459M & 4029& 1400\\
 & 10600 & ACS/FR459M & 4401& 1020\\ 
 & 10600 & ACS/FR459M & 4780& 830\\
 & 10600 & ACS/FR459M & 5165& 885\\
 & 10600 & ACS/FR647M & 5566& 1020\\
 & 10600 & ACS/FR647M & 6244& 1020\\
2010-02-17 & 11629& COS/G140L & 1100--1700 & 10927 \\
2010-02-20 & 11629& COS/G140L & 1100--1700 & 10927 \\
\enddata
\tablecomments{Wavelengths are either the sensitive range of a spectroscopic observation, or the pivot wavelength of a photometric observation.}
\end{deluxetable}

\begin{table*}
\begin{center}
 \centering
  \caption{List of previously published observations and results.}\label{old_phot}
  \begin{tabular}{lcccc}
  \hline
Instrument & Filter & Central Frequency & Flux $F_\nu$ & Reference \\
 & & (10$^{14}$\,Hz) & ($\mu$Jy)\\
\hline\\
HST/WFPC2 & F555W & 5.48 & 0.39(2) & 1\\
HST/FOC & F555W & 5.62 & 0.39(2) & 2\\
HST/FOC & F430W & 7.28 & 0.26(3) & 3\\
HST/FOC & F342W & 8.81 & 0.31(3) & 3\\
HST/FOC & F195W & 12.7 & 0.35(4) & 3\\
BTA/CCD & I & 3.80 & 0.60(6)$^a$ & 4\\
BTA/CCD & R & 4.61 & 0.39(3) & 4\\
BTA/CCD & V & 5.46 & 0.37(4) & 4\\
BTA/CCD & B & 6.84 & 0.32(4) & 4\\
Subaru/Suprime & I & 3.80 & 0.37(4) & 5\\
Subaru/Suprime & R & 4.61 & 0.42(3) & 5\\
Subaru/Suprime & B & 6.84 & 0.31(2) & 5\\
VLT/FORS2 & 4300--9600\,\AA & & & 6\\
\hline\\
\end{tabular}\\
Note: references (1) \citet{1997Msngr..87...43M} (2) \citet{1996ApJ...467..370P}(3) \citet{1997ApJ...489L..75P} (4) \citet{2001A&A...370.1004K} (5) \citet{2006A&A...448..313S} (6) \citet{2007Ap&SS.308..545Z}. Numbers in parentheses indicate the uncertainty in the final digit.

$^a$: The authors note that this measurement was contaminated by a nearby red galaxy.
\end{center}
\end{table*}

Below, we briefly describe the fluxes and spectra derived for each data-set, in order of increasing wavelength. Full details of the data reduction procedures used are given in the Appendix. We encourage interested readers to refer to the Appendix for a thorough account of the specific procedures required for the analysis of the diverse data, in particular  grating spectroscopy (with a slit), slit-less prism spectroscopy, very faint infrared photometry (requiring custom NICMOS darks), ramp filter photometry and the evaluation of calibration performance in the IR-optical regime.

\subsection{STIS FUV grating}

\begin{figure*}
\includegraphics[width=0.7\hsize,angle=90]{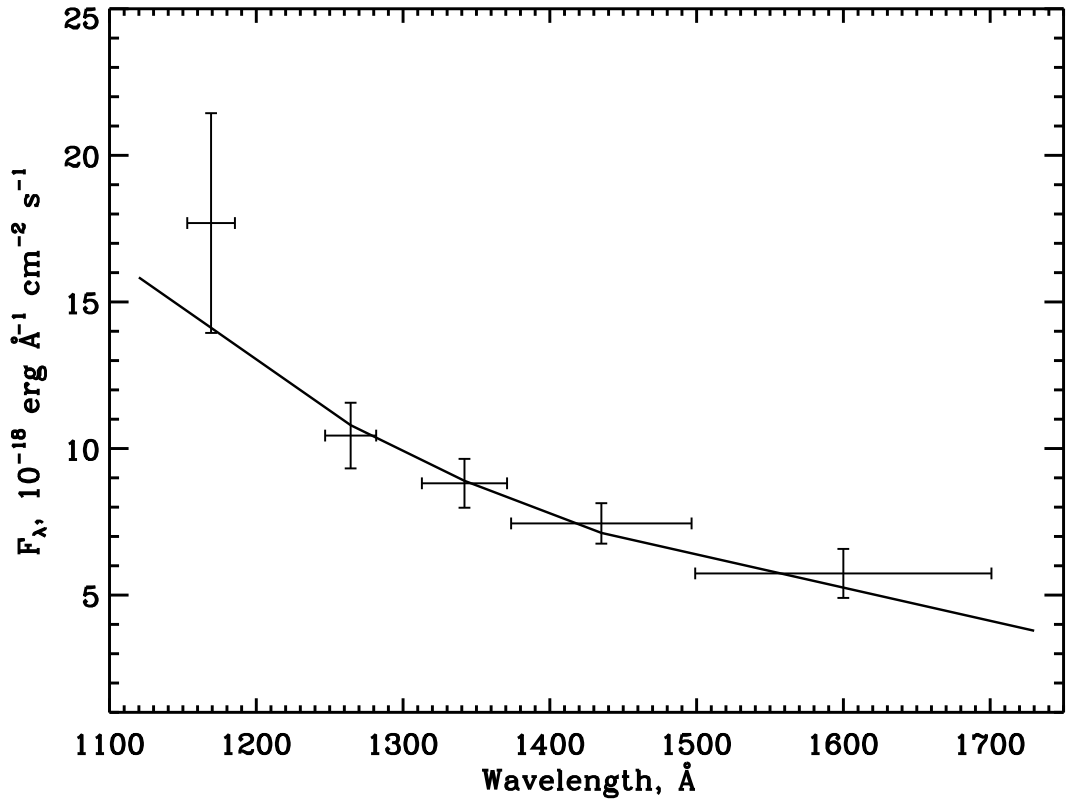}
\caption{STIS FUV spectrum of B0656. The solid line shows an absorbed black-body spectrum with $T=0.663$\,MK and $E(B-V)=0.02$.}\label{raw:fuv:spec}
\end{figure*}

The measured flux values are given in Table
\ref{tab:fuv}, while the spectrum is shown in Figure
\ref{raw:fuv:spec}. The total flux in the 1153--1700 \AA\ range
$(\Delta\lambda=547$\,\AA),
 can be estimated as
$F\simeq \Delta\lambda\, \left(\sum_i \langle F_\lambda\rangle_i
\Delta\lambda_i\right) \left(\sum_i \Delta\lambda_i\right)^{-1}
\simeq (4.28\pm 0.28) \times 10^{-15}$ erg s$^{-1}$ cm$^{-2}$,
corresponding to the luminosity $L_{\rm FUV}=4\pi d^2 F =(4.24\pm
0.28)\times 10^{28} d_{288}^2$ erg s$^{-1}$.

\begin{table*}[!H]
\caption[]{0656 FUV-MAMA total counts and fluxes in $\lambda$-bins \label{tab:fuv}}
\begin{center}
\begin{tabular}{cccc}
\tableline\tableline 
$\lambda$-bin (\AA) & Counts & Background & $f_\lambda$ (10$^{-19}$\,erg/s/cm$^2$/\AA) \\
\tableline
   1153$-$1185 &          48 &      14 &     177(37) \\
      1247$-$1282 &      142 &      33 &    104(11) \\
      1313$-$1371 &      190 &      41 &    88(8) \\
      1372$-$1497 &      223 &      46 &    74(7) \\
      1498$-$1700 &      114 &      25 &    57(8) \\
\tableline
\end{tabular}
\end{center}
\end{table*}

For a typical
 NS radius $R=13$\,km and  distance $d=288$\,pc,
the inferred brightness temperatures are $0.615\pm 0.026$, $0.663\pm
0.028$, $0.715 \pm 0.031$, and $0.832\pm 0.036$ MK,
 for
$E(B-V)=0.01$, 0.02, 0.03, and 0.05, respectively; the corresponding
$\chi_\nu^2$ values are 0.663, 0.642, 0.626, and 0.611, for 4 degrees of freedom (dof).
 An example of
best-fit blackbody spectrum is shown in Figure \ref{raw:fuv:spec},
for a plausible  $E(B-V)=0.02$. The corresponding unabsorbed bolometric
luminosity can be estimated as
 $L_{\rm bol}=1.2\times 10^{29}\, T_5^4 R_{13}^2$
erg s$^{-1}$; for instance, $L_{\rm bol}=(2.3\pm 0.4)\times 10^{32}$
erg s$^{-1}$ for $E(B-V)=0.02$, $R=13$ km and $d=288$ pc.

\subsection{COS spectroscopy}\label{cos_section}

\begin{figure*}
 \centering
\includegraphics[width=0.8\hsize]{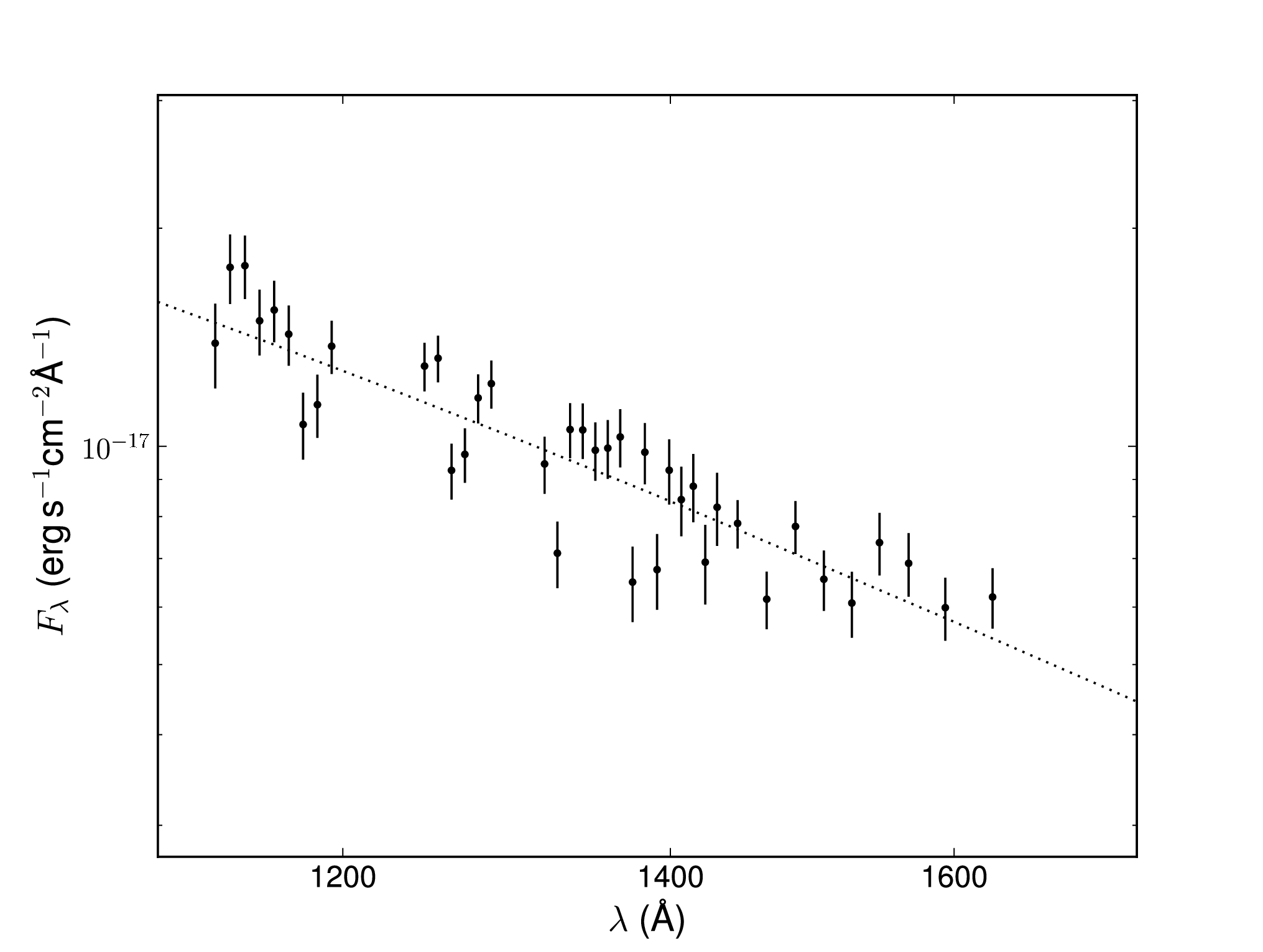}
\caption{COS-FUV spectrum of \psr. The gaps are due to the discarded data strongly affected by geocoronal emission lines (see Section \ref{cos_section}). An absorbed PL fit is plotted, with $F_\lambda = \lambda^{-\alpha_\lambda}\times10^{-0.4A(\lambda)E(B-V)}$, $\alpha_\lambda=3.0\pm0.2$, $E(B-V)=0.03$. }\label{cos_spec}
\end{figure*}

The measured spectrum is shown in Figure \ref{cos_spec} while the spectral fluxes are listed in Table \ref{cos_flux}. An absorbed PL, $F_\lambda\propto\lambda^{-\alpha_\lambda}$, with best-fit $\alpha_\lambda=3.0\pm0.2$ is plotted. The curve follows the overall shape of the spectrum well, but there is substantial scatter resulting in $\chi^2_\nu = 2.1$. The majority of this scatter is in a few outlier points which not appear to form spectral feature(s). We cannot rule out narrow lines, but such lines are not expected for the high temperature and pressure of a NS atmosphere or magnetosphere. We find no obvious reason for the additional scatter, and speculate that it may be due to the imperfections of the calibration or structure in the detector background.

\begin{deluxetable}{ccccc}
\tablecaption{COS FUV total counts and fluxes of B0656 \label{cos_flux}}
\tabletypesize{\footnotesize}
\tablewidth{0in}
\tablehead{
\colhead{$\lambda$ (\AA)} &\colhead{$\delta\lambda$ (\AA)} & \colhead{Counts} & \colhead{Background} & \colhead{$f_\lambda$ (10$^{-19}$\,erg/s/cm$^2$/\AA) }
}
\startdata 
1130.1 & 8.0 & 170 & 70 & 138.8$\pm$18.5 \\
1138.0 & 8.0 & 228 & 87 & 176.6$\pm$19.4 \\
1146.0 & 8.0 & 231 & 74 & 177.5$\pm$17.9 \\
1153.9 & 8.0 & 214 & 69 & 149.0$\pm$15.5 \\
1161.9 & 8.0 & 244 & 81 & 154.2$\pm$15.0 \\
1169.8 & 8.0 & 234 & 70 & 142.8$\pm$13.6 \\
1177.8 & 8.0 & 189 & 57 & 107.2$\pm$11.3 \\
1185.8 & 8.0 & 223 & 73 & 114.2$\pm$11.4 \\
1193.7 & 8.0 & 241 & 51 & 137.5$\pm$11.6 \\
1247.1 & 8.0 & 273 & 55 & 129.0$\pm$9.9 \\
1255.0 & 8.0 & 274 & 45 & 132.3$\pm$9.8 \\
1263.0 & 8.0 & 203 & 39 & 92.7$\pm$8.2 \\
1270.9 & 8.0 & 222 & 48 & 97.5$\pm$8.4 \\
1278.9 & 8.0 & 251 & 43 & 116.6$\pm$9.1 \\
1286.9 & 8.0 & 260 & 44 & 122.0$\pm$9.3 \\
1319.5 & 8.0 & 195 & 39 & 94.6$\pm$8.5 \\
1327.5 & 8.0 & 144 & 29 & 71.2$\pm$7.5 \\
1335.5 & 8.0 & 205 & 39 & 105.5$\pm$9.2 \\
1343.4 & 8.0 & 198 & 36 & 105.3$\pm$9.3 \\
1351.4 & 8.0 & 187 & 38 & 98.8$\pm$9.1 \\
1359.4 & 8.0 & 184 & 38 & 99.4$\pm$9.3 \\
1367.4 & 8.0 & 182 & 34 & 103.0$\pm$9.5 \\
1375.3 & 8.0 & 120 & 28 & 65.0$\pm$7.7 \\
1383.3 & 8.0 & 171 & 36 & 98.2$\pm$9.5 \\
1391.3 & 8.0 & 121 & 30 & 67.6$\pm$8.1 \\
1399.2 & 8.0 & 161 & 39 & 92.7$\pm$9.6 \\
1407.2 & 8.0 & 147 & 38 & 84.4$\pm$9.3 \\
1415.2 & 8.0 & 146 & 35 & 88.1$\pm$9.5 \\
1423.2 & 8.0 & 121 & 36 & 69.2$\pm$8.7 \\
1431.1 & 8.0 & 136 & 37 & 82.4$\pm$9.5 \\
1445.0 & 19.9 & 308 & 83 & 78.3$\pm$6.0 \\
1464.9 & 19.9 & 249 & 83 & 61.5$\pm$5.6 \\
1484.9 & 19.9 & 286 & 92 & 77.5$\pm$6.5 \\
1504.9 & 20.0 & 236 & 84 & 65.6$\pm$6.3 \\
1524.8 & 20.0 & 210 & 80 & 60.8$\pm$6.3 \\
1544.8 & 20.0 & 234 & 89 & 73.6$\pm$7.3 \\
1566.1 & 22.8 & 232 & 91 & 69.0$\pm$6.9 \\
1593.4 & 32.0 & 293 & 141 & 59.9$\pm$6.0 \\
1629.2 & 40.0 & 343 & 173 & 62.0$\pm$5.9 \\
\enddata
\end{deluxetable}

\subsection{STIS NUV prism}

\begin{figure*}[!H]
 \centering
\includegraphics[width=0.7\hsize,angle=90]{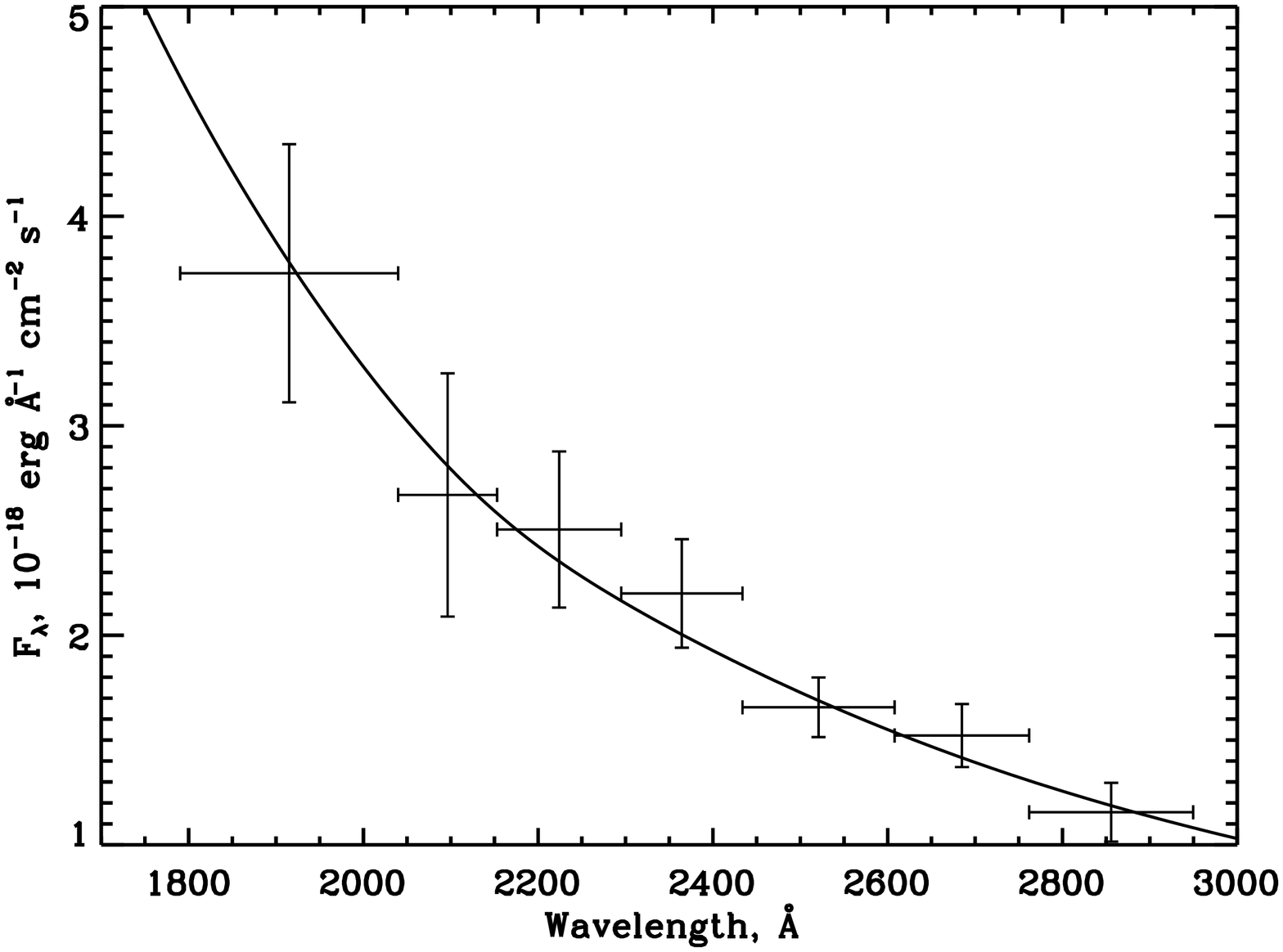}
\caption{ The measured (absorbed) NUV-MAMA spectrum of B0656. The
 curve
 shows the best-fit absorbed PL model with $\alpha_{\lambda}=-3.01$
for $E(B-V)=0.02$. }\label{raw:nuv:spec}
\end{figure*}

 The
measured flux values are given in Table \ref{tab:nuv}, while the
spectrum is shown in Figure \ref{raw:nuv:spec}. The total flux in
the 1790--2950 \AA\ range $(\Delta\lambda=1160$\,\AA),
 can be estimated as
$
(2.63\pm 0.38) \times 10^{-15}$ erg s$^{-1}$ cm$^{-2}$,
corresponding to the luminosity $L_{\rm NUV}=4\pi d^2 F =(2.61\pm
0.38)\times 10^{28} d_{288}^2$ erg s$^{-1}$. The obtained average
flux is slightly larger, but compatible with the flux measured by Shibanov et al.\ (2005)
in somewhat wider wavelength interval ($1700-3400$ \AA).

We then fit the NUV spectrum with the absorbed power-law model,
$F_{\lambda}= F_{2500}\,(\lambda/2500\,$ 
\AA)$^{\alpha_{\lambda}}\times 10^{-0.4 A(\lambda)\, E(B-V)}$. For
plausible values $E(B-V)=0.01$, 0.02, 0.03 and 0.05, we find the
power-law indices $\alpha_\lambda = -2.92\pm 0.41$, $-3.01\pm 0.41$,
$-3.09\pm 0.41$ and $-3.24\pm 0.41$, and the normalization $F_{2500}
= 1.85\pm0.08$, $1.97\pm0.09$, $2.11\pm 0.10$ and $2.41\pm 0.11
\times 10^{-18}$ erg cm$^{-2}$ s$^{-1}$ \AA$^{-1}$, respectively; the corresponding $\chi_\nu^2$ values are
1.27, 1.38, 1.51, and 1.87, for 5 dof. The
obtained spectral slopes are systematically smaller than those
derived from our fits to the FUV spectrum. The flattening of the
spectrum at longer wavelengths can be attributed to a larger
relative contribution of the non-thermal component. The measured
slope of the NUV spectrum for $E(B-V)=0.03$,
$\alpha_\nu\simeq1.1\pm0.4$ ($\alpha_{\nu}=-\alpha_{\lambda}-2$), is marginally compatible with the slope
$\alpha_\nu=0.35\pm0.5$ found by Shibanov et al.\ (2005). The difference can be attributed to the different
choice of the spectral bins and narrower wavelength range that we
used. 

\begin{table*}[!H]
\caption[]{B0656 NUV-MAMA total counts and fluxes in nine $\lambda$-bins
($+4$ pixels shift in the dispersion direction is applied) \label{tab:nuv}}
\begin{center}
\begin{tabular}{cccc}
\tableline\tableline 
$\lambda$-bin (\AA) & Counts & Background& $f_\lambda$ ($10^{-19}$\,erg/s/cm$^2$/\AA) \\
\tableline
      1790$-$2040 &      2594 &      2316 &  37(6) \\
      2040$-$2153 &      1781 &      1606 &  27(6) \\
      2153$-$2295 &      1889 &      1634 &  25(4) \\
      2295$-$2434 &      1514 &      1251 &  22(3) \\
      2434$-$2608 &      1590 &      1274 &  16.5(14) \\
      2608$-$2762 &      1173 &      909 &   15.2(15)\\
      2762$-$2950 &      1136 &      913 &   11.6(14) \\
\tableline
\end{tabular}
\end{center}
\end{table*}

\subsection{ACS/WFC Ramp Filter Photometry}

\begin{deluxetable}{cc}
\tablecaption{ACS/WFC ramp filter fluxes of B0656 \label{ramp_flux}}
\tabletypesize{\footnotesize}
\tablewidth{0in}
\tablehead{
\colhead{$\lambda$ (\AA)} & \colhead{$f_\nu$ ($\mu$Jy) }
}
\startdata 
4049 & 0.31$\pm$0.3\\
4401 & 0.46$\pm$0.3\\
4780 & 0.38$\pm$0.3\\
5165 & 0.35$\pm$0.2\\
5566 & 0.45$\pm$0.3\\
6244 & 0.39$\pm$0.3\\
6822 & 0.49$\pm$0.2\\
7393 & 0.40$\pm$0.3\\
7956 & 0.48$\pm$0.3\\
8751 & 0.46$\pm$0.3\\
9491 & 0.35$\pm$0.5
\enddata
\end{deluxetable}

The photometric fluxes for three ramp filters and a total of 11 central wavelength positions are listed in Table \ref{ramp_flux} and shown in Figure \ref{spectrum}. Given the non-standard nature of the ramp filter photometry, we performed a specific check on our calibration in Section \ref{check}. We find the pipeline calibration  satisfactory.

\subsection{NICMOS photometry}

We find $F_\nu({\rm F110W}) = 0.385\pm0.030$\,$\mu$Jy; 
$F_\nu({\rm F160W}) = 0.551\pm0.030$\,$\mu$Jy; 
$F_\nu({\rm F187W}) = 0.660\pm0.025$\,$\mu$Jy (uncertainties are 1-$\sigma$ and statistical only). Due to the non-standard processing steps and calibration involved, we conservatively estimate the systematic uncertainty to be  about 10\% in each case (see Section \ref{check}). Our flux values are only marginally consistent with  those reported by \citet{2001A&A...370.1004K}. We believe that by using the most up-to-date calibration, our values should be the most accurate. We verify our calibration in Section \ref{check}.

\section{Results and discussion}

\subsection{IR-FUV spectrum}\label{iruv}
\begin{figure*}
\begin{center}
\includegraphics[width=0.99\hsize]{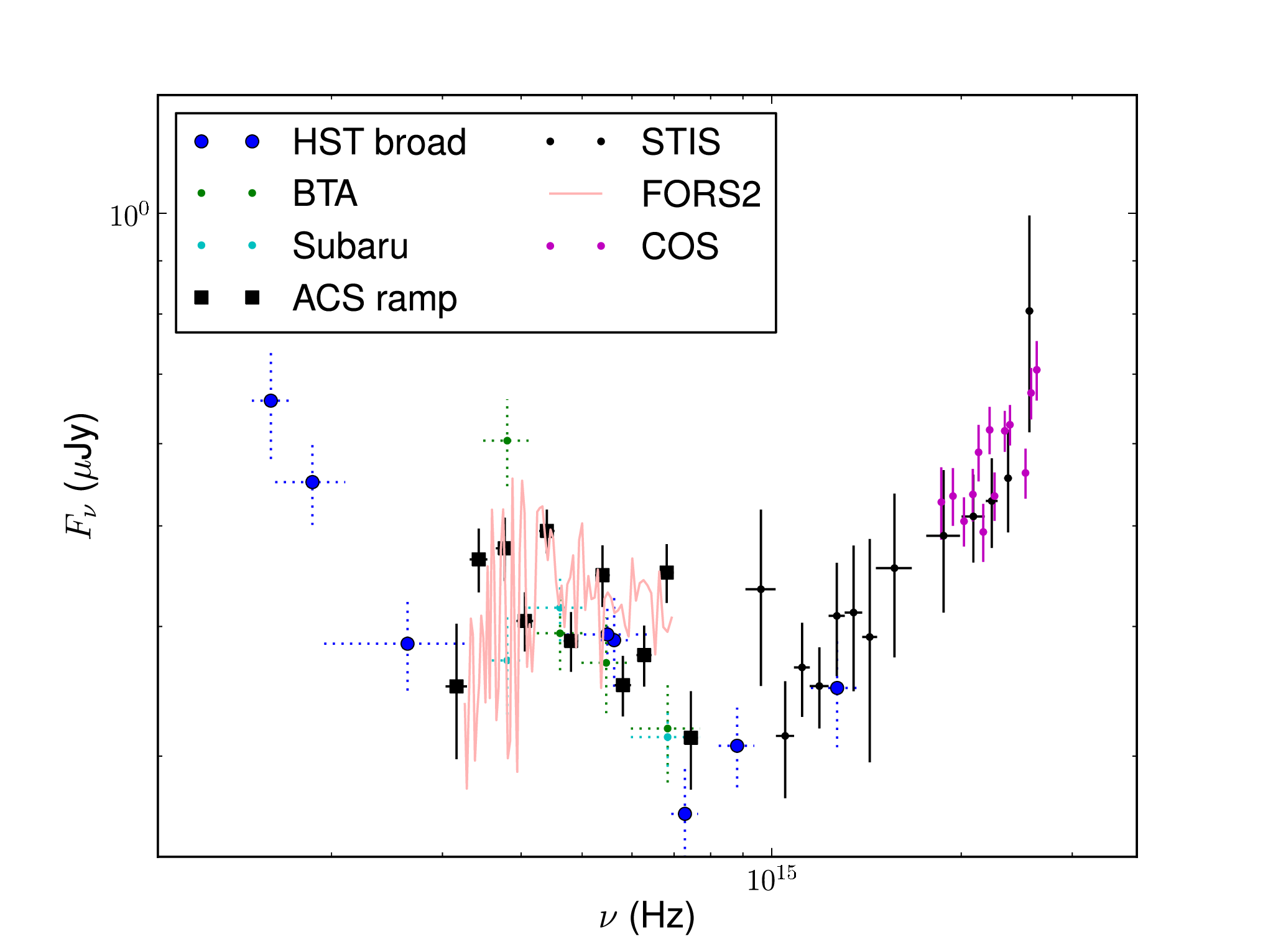}
\caption{Spectrum of \psr\ (as observed). The data plotted are a combination of fluxes in the literature (Table \ref{old_phot}) and our own work (Table \ref{log}).}\label{spectrum}
\end{center}
\end{figure*}

The final spectrum shown in Figure \ref{spectrum} includes both the fluxes/spectra calculated in this work and those from the literature (see Table \ref{old_phot}). This is, to our knowledge, the most complete IR-UV spectrum of a non-recycled pulsar.

\subsubsection{Qualitative spectral shape}\label{fit_text}
Clearly, the spectrum of B0656 shown  in Figure \ref{spectrum} does not fit to a simple absorbed PL model. The spectrum appears to be more complex, with some structure. Therefore we begin by making some qualitative statements.

The upturn of the spectrum in the NUV range is consistent with a contribution by a
thermal, Rayleigh-Jeans (R-J; $F_\nu\propto\nu^2$) component that is most prominent in FUV
 and also consistent with the extrapolation of the X-ray TS component to the optical (see also \citealt{2007Ap&SS.308..287K}). However, even with the thermal component added, any simple PL
fit to the optical spectrum of B0656 leaves large residuals. In particular, there is a deep minimum in the spectral flux around $\nu=3\times10^{14}$\,Hz ($\lambda=1\,\mu$m), with steep rises to either side (see Figure \ref{spectrum}). 

\begin{figure*}
\begin{center}
\includegraphics[width=0.59\hsize]{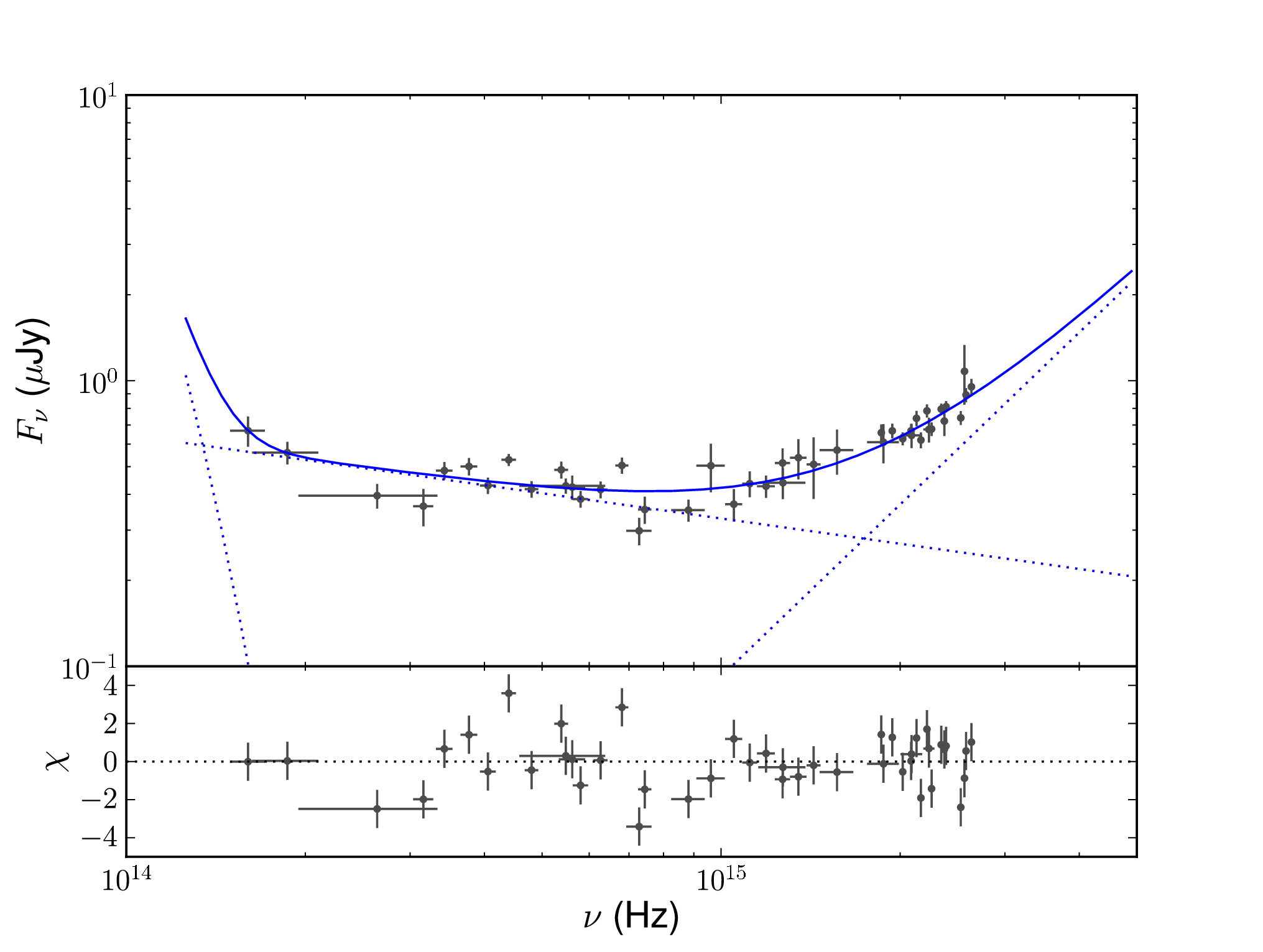}\vspace{-0.7cm}
\includegraphics[width=0.59\hsize]{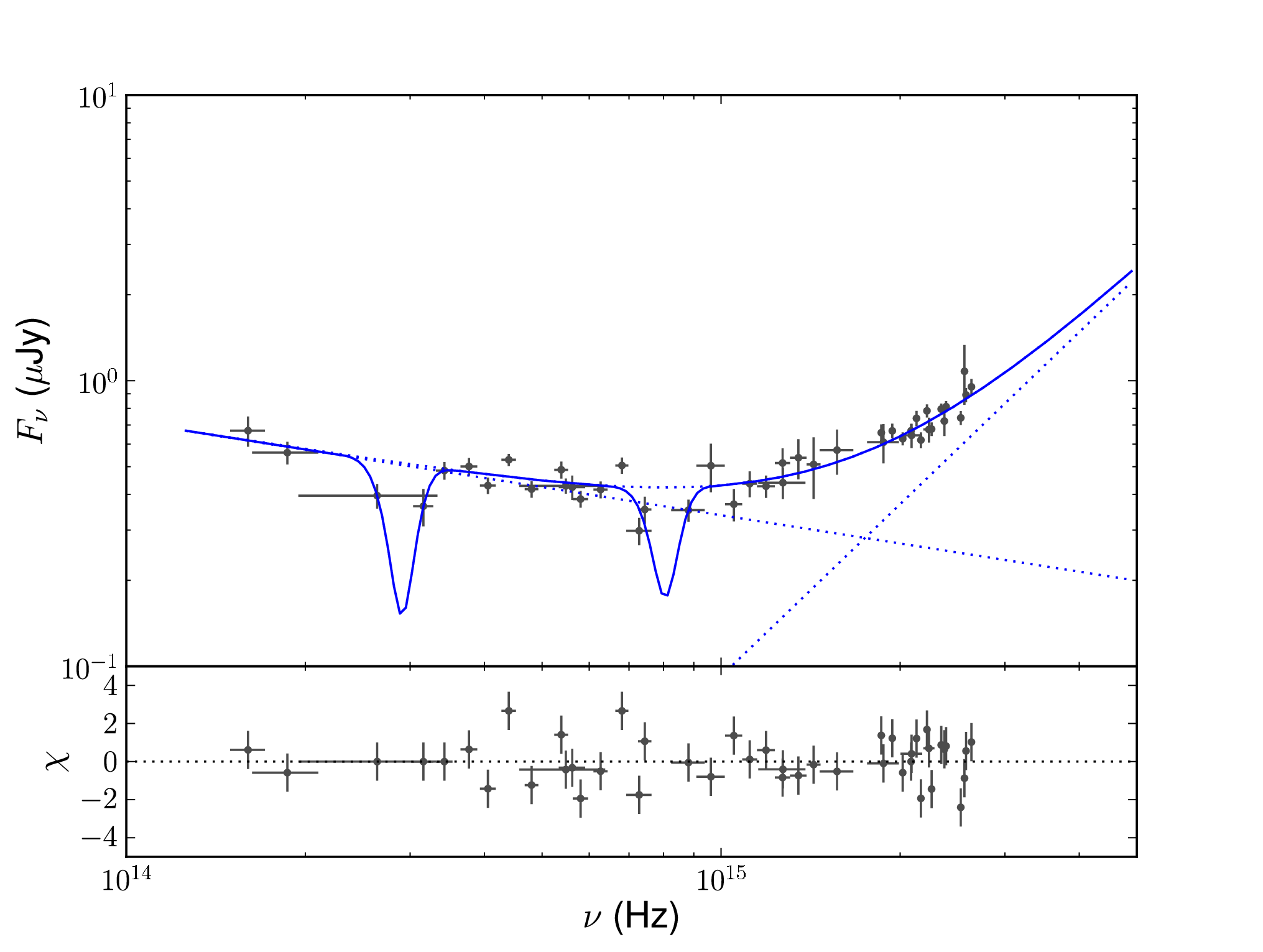}\vspace{-0.7cm}
\includegraphics[width=0.59\hsize]{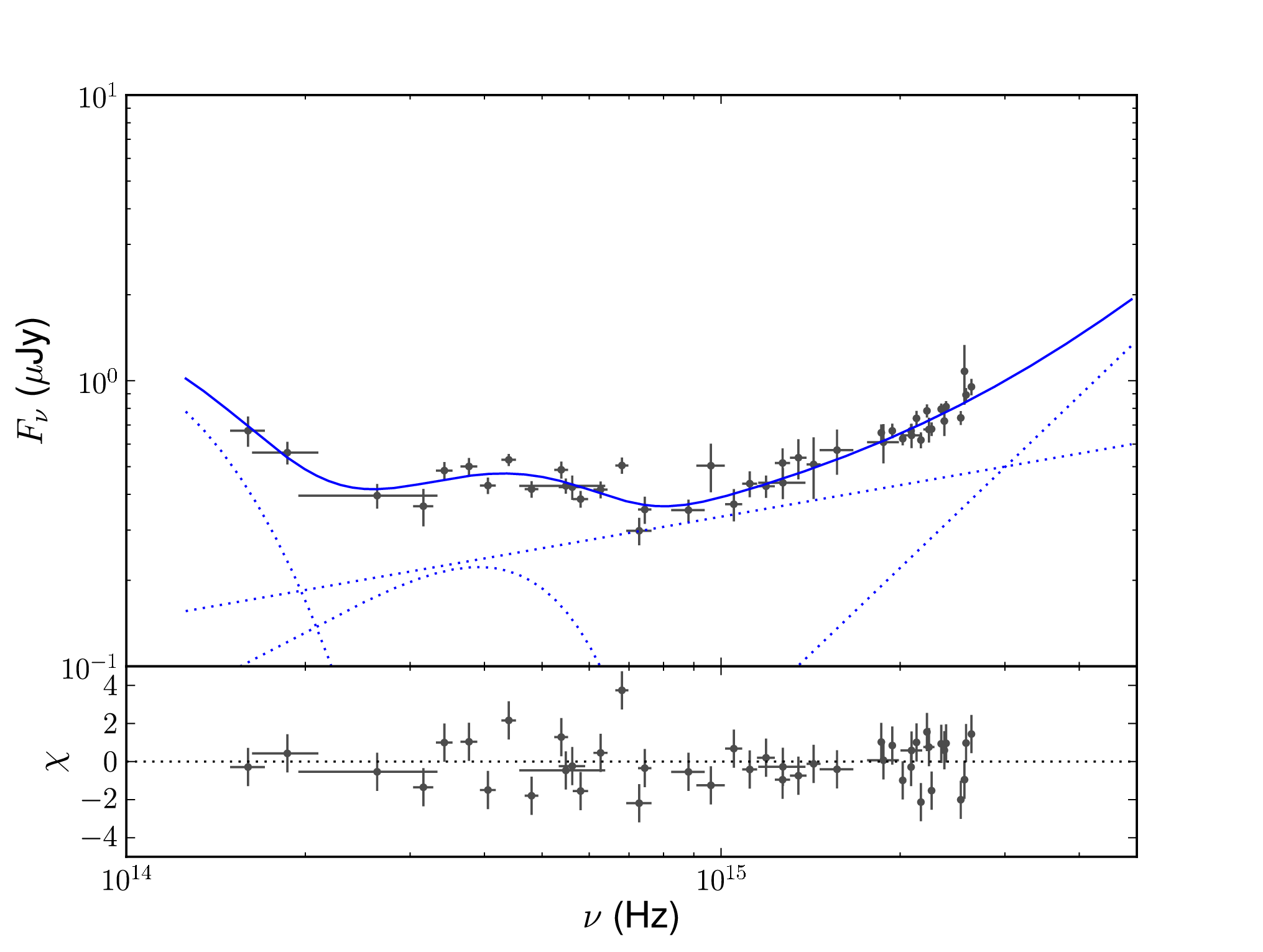}
\caption{Spectral fits to the IR-UV spectrum of \psr: continuum-only model (top), absorption lines (middle) or broad emission line (bottom) components.}\label{fits}
\end{center}
\end{figure*}

The  complex spectral shape can be interpreted in several ways. For instance, the steep rise at the lowest frequencies could be attributed to a cool, $T\sim1000$\,K, thermal (disk) component. Alternatively, the spectrum is very suggestive of {\em spectral features}, either two absorption lines or a very broad emission  line. We have fitted models corresponding to these three scenarios with the best-fit parameters listed in Table \ref{fit_table}. The corresponding models and residuals are shown in Figure \ref{fits}. We note that none of the fits is statistically good ($\chi^2_\nu>1.5$), but these fits do not show systematic structure in the residuals. Of the three, the absorption model, with lines at (2.91$\pm0.03$) and (8.05$\pm0.09)\times10^{14}$\,Hz ($\lambda=1.030\pm$0.011 and 0.372$\pm$0.004\,$\mu$m), is statistically preferred.

\begin{deluxetable}{ccccccc}
\tablecaption{Simple fits to the IR-UV spectrum \label{fit_table}}
\tabletypesize{\footnotesize}
\tablewidth{0in}
\tablehead{
\colhead{Model} & \colhead{$T_{\rm cold}$ (K)} & \colhead{$\alpha_\nu$} & \colhead{$N_{\rm PL}$\tablenotemark{a}}& \colhead{$f_{\rm hot}$\tablenotemark{b}} & \colhead{$\nu_{\rm line}$ (10$^{14}$\,Hz)} & \colhead{$\chi^2/dof$}
}
\startdata 
Continua & 500 & $-$0.29(6) &0.404(8)& 0.37(2) & \ldots& 86/40\\
Absoption& \ldots& $-$0.32(5) &0.424(9)& 0.37(2) & 2.91(3), 8.05(9) & 58/36\\
Emission & 1200(400) & 0.4 &0.26(6)& 0.22(13) & 3.9(5) & 66/37 \\
\enddata
\tablecomments{Numbers in parentheses indicate uncertainty in the final digit(s). Where no uncertainty is given, the value is very poorly defined ($<1\sigma$).}
\tablenotetext{a}{Power-law ($F_\nu = N_{\rm PL} \nu^{-\alpha_\nu}$) normalization at $\nu=5\times10^{14}$\,Hz, in $\mu$Jy.}
\tablenotetext{b}{Flux density of the Rayleigh-Jeans power-law at $\nu=2\times10^{15}$\,Hz, in $\mu$Jy.}
\end{deluxetable}

\subsubsection{Possible interpretations}
Several possible interpretations for spectral features can be considered. As an example, we illustrate the apparently straightforward case of {\em cyclotron} emission (or absorption). Given the current uncertainties of the optical flux, we could speculate that we are observing cyclotron lines from a cool plasma. The NS magnetosphere has a dipole magnetic field $B\propto(R_{NS}/r)^{3}$ with surface strength $B_{\rm dipole}\sim 5\times 10^{12}$\,G . An electron cyclotron line centred at $\lambda\approx1\,\mu$m would require magnetic field $B\sim2\times10^8$ G, which would place the line forming region at $d\approx300R_{NS}$ from the NS surface but still well within the light cylinder radius ($R_{\rm LC} = 1.15\times10^5 R_{\rm NS}$). 

A deep dip at 1\,$\mu$m ($\nu=3\times10^{14}$\,Hz; Figure \ref{spectrum}) could signify a strong absorption line, and so suggest the hypothesis of an absorbing cyclotron-resonance layer which could be composed of $e^{+}/e^{-}$ \citep{1997ApJ...491..296R,1998ApJ...498..373W,2003astro.ph.10777R}. Such a layer can significantly reprocess/change thermal radiation from the hot NS surface. If confirmed by more sensitive observations of B0656,  quantitative measurements (centroid, line width, phase dependence) will open a new window to probe the structure of the pulsar magnetosphere.  

Alternatively, there may be a {\em cold thermal} component in the infra-red (a steepening rise towards the red) which could be indicative of a passive post-supernova fall-back disk \citep{1989ApJ...346..847C} for which we estimate a temperature of about 500-1500\,K.  This would suggest a link to magnetars, where such fall-back disks have already been suggested \citep{2006Natur.440..772W}, and support recent models that predict the persistence of fall-back disks to much later epochs than recently thought \citep{2011MNRAS.416..893D}.

\subsection{Broad-band multi-wavelength spectrum}
Let us consider the NIR-UV spectrum of B0656 in a broader, multi-wavelength context. B0656 has been extensively studied in X-rays (e.g., \citealt{2002nsps.conf..273P,2005ApJ...623.1051D}), and has also been detected as a $\gamma$-ray source by {\sl Fermi} LAT with high significance \citep{2010ApJS..187..460A}. 

We downloaded  {\sl Chandra} ObsID 2800 (25\,ks  exposure with ACIS in continuous clocking [CC] mode; see also \citealt{2002nsps.conf..273P}) and also the spectrum of ObsID 130 (38\,ks exposure with the HRC and low-energy transmission grating [LETG]; see  \citealt{2002ApJ...574..377M}). The data were reduced using standard procedures. We convert the counts spectra to fluxes by dividing the count rate in each bin by the calibrated effective area in that bin (sometimes called ARF), after convolving with the response function (sometimes called RMF). Although such `fluxing' can produce an incorrect  spectral shape at sharp detector sensitivity edges and spectral features\footnote{X-ray detectors are prone to `redistribution', the possibility that a photon produces a signal corresponding to a different energy.}, the deviations are confined to relatively narrow energy ranges, and the spectrum obtained in this way is independent of any pre-supposed spectral model. We also used the fluxes from the second {\sl Fermi} catalog (2FGL;  \citealt{2011arXiv1108.1435T}).

\begin{figure*}
\begin{center}
\includegraphics[width=0.7\hsize]{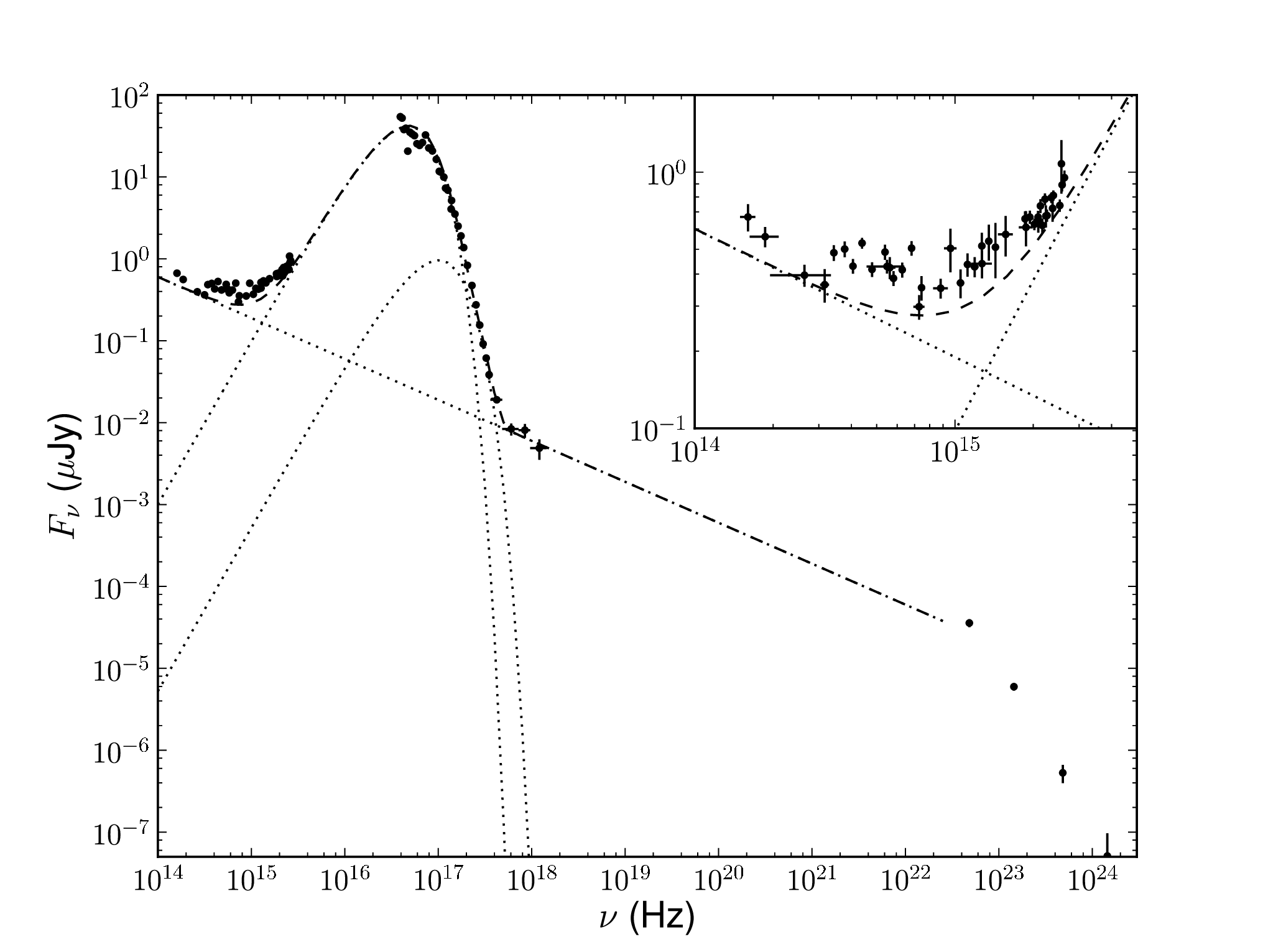}
\caption{Multiwavelength spectrum of B0656, including the NIR-UV data, {\sl Chandra} X-ray fluxed spectrum (this work) and $\gamma$-ray fluxes from the 2FGL catalog. We also sketch the fit to the X-ray spectrum by \citet{2004MmSAI..75..458Z}: two black-bodies with temperatures $T=0.82,1.72$\,MK and a power-law with photon index $\Gamma=1.5$. The X-ray absorbing column is $N_{\rm H}=1\times10^{20}$\,cm$^{-2}$, and the optical/UV reddening $E(B-V)=0.03$.}\label{MWspec}
\end{center}
\end{figure*}

In Figure \ref{MWspec} we show these  data along with the NIR-UV fluxes.  Note, however, that the extinction (and optical/UV reddening) is rather uncertain, and has a particularly strong effect at the softest X-ray energies ($E\sim100$\,eV).

We have plotted in Figure \ref{MWspec} the best-fit X-ray model from \citet{2002nsps.conf..273P}, fitted to the same {\sl Chandra} data\footnote{\citet{2005ApJ...623.1051D} fitted the same model to {\sl XMM-Newton} data, and found fit parameters formally inconsistent with the ones obtained from the {\sl Chandra} data; whereas  \citep{2004MmSAI..75..458Z} found consistent parameters. We  note that even within the {\sl XMM-Newton} data there are obvious discrepancies between the spectra from two different instrument modes for $E<0.5$\,keV (see Figure 2 from \citealt{2005ApJ...623.1051D}) and therefore the best-fit uncertainties have likely been underestimated.}.
The extrapolation of the X-ray model, the sum of two black-bodies and a power-law, is shown in the inset of Figure \ref{MWspec}. Surprisingly, both components fall close to the UV-optical spectrum (Section \ref{iruv}). Extrapolating the X-ray fits into the NIR-UV exacerbates the uncertainties in the model flux: a small uncertainty in any given parameter translates into a large uncertainty in the predicted flux in the NIR-UV regime. It is natural to assume that the Rayleigh-Jeans spectral component is a continuation of the cooler X-ray black-body; for a temperature of $T=0.82$\,MK, its radius would be $R\approx10$\,km (as observed at infinity) to fit the FUV, with a 15\% uncertainty on $T$ for the reddening range 0.01$<E(B-V)<$0.05.
The power-law component in the optical could, in principle, be a continuation of non-thermal power-law emission seen in X-rays (but would not be sufficient to describe the data without the features or a cool disk; see Section \ref{fit_text}). 

The extrapolation of the best-fit PL also happens to fall close to the {\sl Fermi} $\gamma$-rays fluxes, hinting that a single very broad PL with  photon index $\Gamma_{\rm MW}=1.50\pm0.05$ could in principle describe the entire non-thermal spectrum from optical to $\gamma$-rays. In order to match the GeV fluxes, the PL should of course be modified by a high-energy cut-off, normally seen for $\gamma$-ray pulsars \citep{2010ApJS..187..460A}. The high pulsed fraction ($>$50\%) in the {\sl Fermi} LAT  light-curve \citep{2010ApJS..187..460A} and also in the optical \citep{2003ApJ...597.1049K} confirms that both are magnetospheric. 

\subsection{Comparison to other pulsars}

Less than 1\% of radio pulsars have been detected in the NIR-optical-UV spectral window to date. Of those all have been detected in X-rays but only some in $\gamma$-rays (Table \ref{irouv_table}). We note that this sample includes members that are young and energetic, middle-age pulsars such as B0656+14, and one millisecond (recycled) pulsar (MSPs). No isolated (i.e., non-accreting) pulsar has yet been discovered in the IR-UV range without having previously been detected in the radio and X-rays. 

We retrieved the IR/optical/UV fluxes and X-ray spectra of the  pulsars in Table \ref{irouv_table},  using our previous work (e.g., Durant et al., 2011; \citealt{2001ApJ...552L.129P}; \citealt{2004MmSAI..75..458Z}), X-ray spectral products downloaded from the Xassist project\footnote{\tt http://xassist.pha.jhu.edu/zope/xassist}, or, for the case of B1951+32, by re-processing data from the archives. Note  that the low-significance optical detection of B1951+32 \citep{2002A&A...395..845B,2004ApJ...610L..33M} has not been confirmed. The GeV fluxes were taken from the 2FLG catalog \citep{2011arXiv1108.1435T}.

\begin{figure*}
\includegraphics[width=0.9\hsize,height=0.75\hsize]{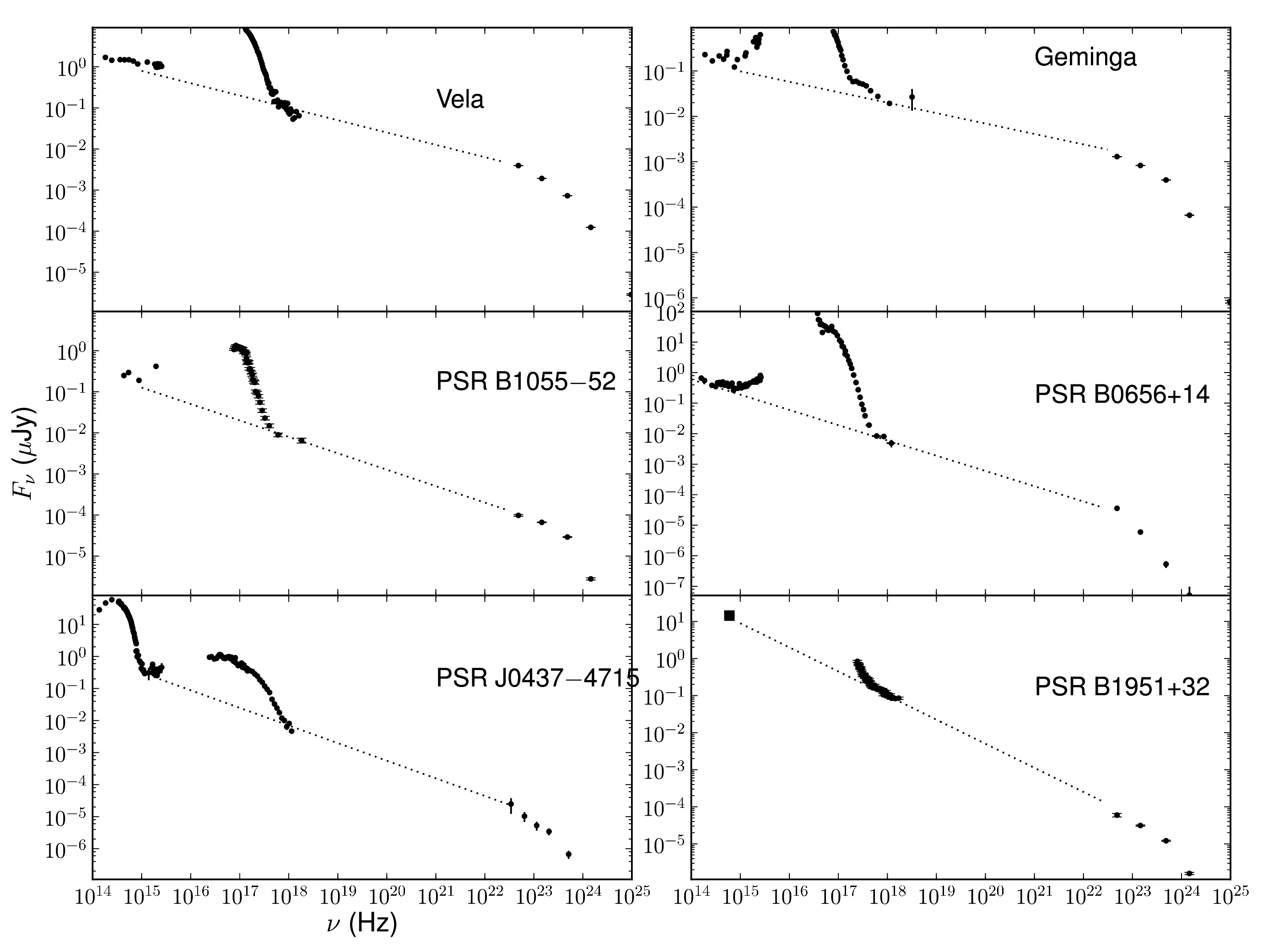}
\caption{NIR/optical, X-ray and GeV spectra of pulsars detected in the NIR-UV window. Dotted lines are PLs drawn to connect the optical and $\gamma$-ray points (see text for discussion). Note that for J0437$-$4715, the lowest frequencies are dominated by its WD companion (see Durant et al. 2011). }\label{PSRs}
\end{figure*}

\begin{figure*}
\includegraphics[width=0.9\hsize,height=0.75\hsize]{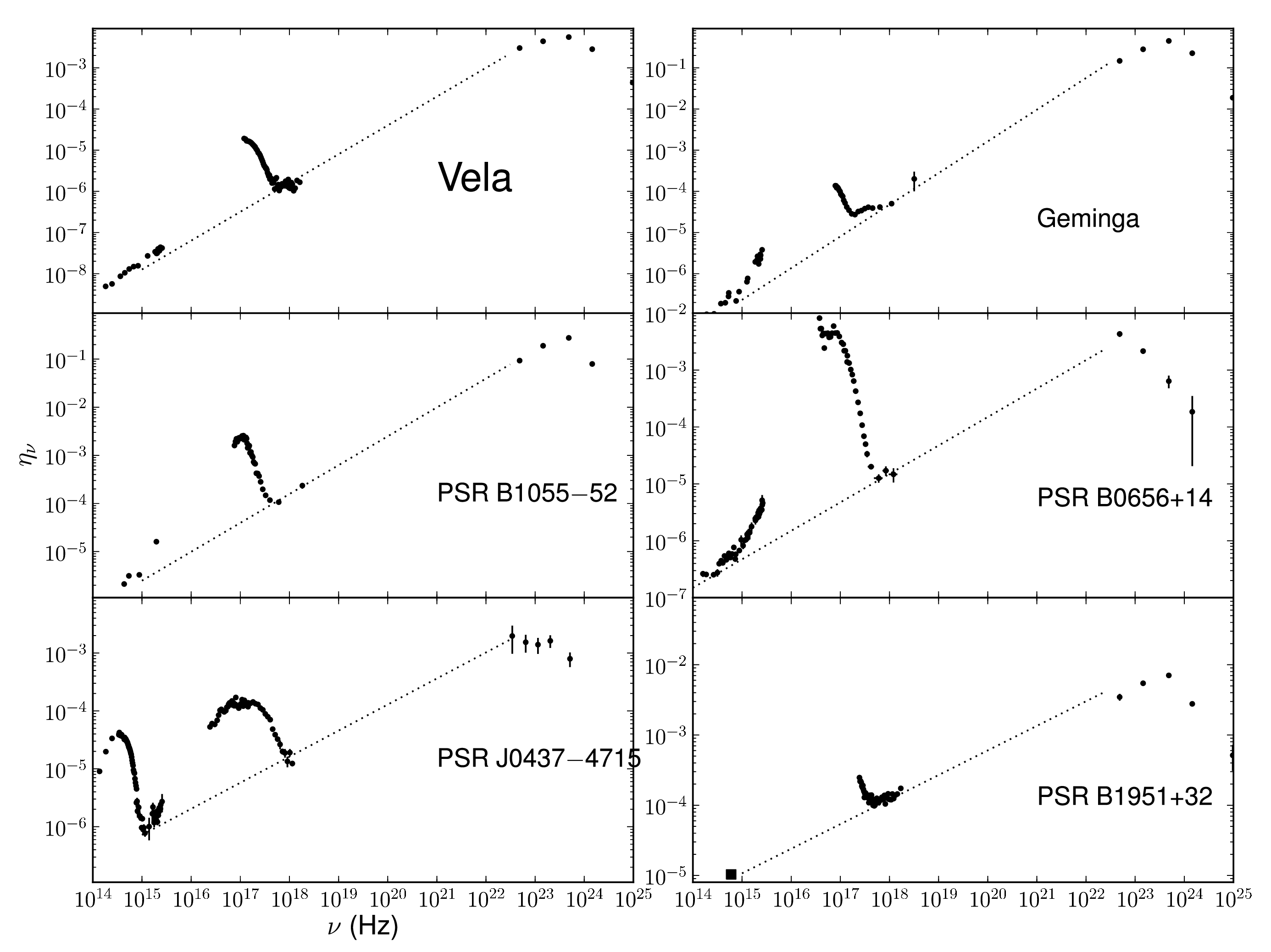}
\caption{As Figure \ref{PSRs}, but in terms of $\eta_\nu \equiv 4\pi d^2\nu F_\nu/\dot{E}$}\label{eta}
\end{figure*}

In Figure \ref{PSRs} we show the IR-GeV spectra of the pulsars in Table \ref{irouv_table}, except for the Crab (see \citealt{2001A&A...378..918K}). Figure \ref {eta} shows the same data, but in terms of `spectral efficiency', $\eta_\nu \equiv 4\pi d^2\nu F_\nu/\dot{E}$. We have sketched PL lines which connect the $\gamma$-ray fluxes with the optical for each spectrum and find that in each case the line also passes through the high-energy, non-thermal part of the X-ray spectrum. This suggests that each pulsar's non-thermal spectrum can be empirically described by a single PL stretching across many orders of magnitude in frequency, with a break or cut-off at GeV energies. Note, that \citet{2010ApJ...712.1209A} have already considered whether the $\gamma$-ray and X-ray non-thermal spectra may be consistent, and conclude that they are not (see their Figure 5); however, there is no discussion there of the uncertainties in the fitted parameters or of the effect of the choice of model function (e.g., atmosphere models for X-rays and broken power-laws in $\gamma$-rays).
In Table \ref{irouv_table} we give the appropriate spectral indices of the lines we have sketched (these are not true fits to the data).
The range of the photon index, $1.2\leq\Gamma_{\rm MW}\leq1.65$, is remarkably narrow. We note that the scales in the Figure are logarithmic, and there may well be additional structure on top of the speculative underlying PL component (as detected in B0656, see Figure \ref{MWspec}), but surprisingly it appears that a remarkably simple description  can be applied for a number of pulsars. We note that such a simple description is not universal, as it is not appropriate for very young pulsars such as the Crab\footnote{The recent high-energy spectrum of the Crab is  shown in the recent poster {\tt confluence.slac.stanford.edu/\\download/attachments/102860834/\\KUIPER\_PSR\_S2.N19\_POSTER.pdf}}.


\begin{deluxetable}{cccccccccccccc}
\tablecaption{Pulsars with IR-UV and GeV detections \label{irouv_table}}
\tabletypesize{\footnotesize}
\tablewidth{0in} 
\rotate
\tablehead{
\colhead{Pulsar} & \colhead{$d$}&\colhead{$\log\tau$} & \colhead{$\log B_0$} & \colhead{$\log B_{\rm LC}$} & \colhead{$\log\dot{E}$} & \multicolumn{6}{c}{ $\log\eta$}  & \colhead{$E_{\rm cut}$} & \colhead{$\Gamma_{\rm MW}$}\\
&\colhead{(kpc)}&&&&& \colhead{NIR} & \colhead{Optical} & \colhead{UV} & \colhead{X$_{\rm therm}$} &\colhead{ X$_{\rm PL}$ } & \colhead{$\gamma$-ray} 
} 
\startdata 
Crab & 2.0(5)&3.09 & 12.58 & 5.99 & 38.7 &$-$5.3&$-$4.8&$-$4.2&\dots&$-$3.9&$-$3&5.8(5)& \dots\\
Vela & 0.287(18)&4.05 & 12.53 & 4.65 & 36.8 &$-$8.3&$-$7.9&$-$7.5&$-$4.6&$-$5.8&$-$2.0&3.2(1)& 1.30\\ 
Geminga &0.25(9)& 5.53& 12.21 & 3.06 &34.5 &$-$7.0&$-$6.7&$-$5.7&$-$5.3&$-$4.3&$-$0.1&1.90(5)& 1.20\\
B0656+14 &0.29(3)&5.05& 12.67 & 2.88 & 34.6 &$-$6.6&$-$6.3&$-$5.6&$-$2.8&$-$4.8&$-$2.0&0.7(5)& 1.50\\
B1055$-$52&0.35(15)&5.73&12.03 & 3.12 & 34.0 &\dots&$-$6.3&$-$5.4&$-$3.4&$-$4.5&$-$0.85&1.3(1)& 1.40\\
B1951+32 &2.0(5)&5.03& 11.69 & 4.87 & 36.6 &\dots&$-$5&\dots&$-$4&$-$3.9&$-$1.7&4.5(1.2)& 1.65\\
J0437$-$4715&0.156(1)&9.83&8.45& 4.45 & 34.1 &\dots&\dots&$-$5.8&$-$4.1&$-$4.8&$-$1.7&1.3(7)& 1.55\\
\enddata
\vspace{-0.3in}
\tablecomments{All values in typical standard units, age $\tau$ in yr, magnetic field $B$ in G, spin-down luminosity $\dot{E}$ in erg/s, and high-energy cut-off $E_{\rm cut}$ in GeV. The efficiency $\eta$ is the fraction of $\dot{E}$ seen in each spectral window. For the efficiencies, we evaluate $\nu L_\nu/\dot{E}$ at 2$\times10^{14}$,  5$\times10^{14}$,  2$\times10^{15}$, and 1$\times10^{18}$\,Hz, for IR, Optical, UV, and X$_{\rm PL}$ respectively. For X$_{\rm therm}$ we calculate the efficiency using  the bolometric luminosity of the hottest thermal component seen, and for the $\gamma$ efficiency we use the 100\,MeV--100\,GeV integrated luminosites in \cite{2010ApJS..187..460A}. The values of $\tau$, $B$ and $\dot{E}$ have been corrected for the effect of proper motion, important for J0437$-$4715. $\Gamma_{\rm MW}$ is PL photon index that connects the IR, X-rays and $\gamma$-rays, with typical uncertainty around 0.05 (Figure \ref{PSRs}).}
\end{deluxetable}

Recently \citet{2011MNRAS.415..867D} presented evidence for steeply-rising infra-red excesses in the Geminga and Vela pulsars (the former only a marginal detection). It may be that this is a common feature of middle-aged pulsars. For Geminga, which has a similar spectrum, luminosity and distance to \psr, they conclude that the IR excess (if it is real) can be successfully modeled with a cool irradiated disk truncated at an inner radius comparable to the light cylinder $R_{LC} = 0.016 R_\odot$. In the case of \psr, we do not have good enough data to attempt to model disk-like emission, and {\sl Spitzer} observations cannot resolve the pulsar from the brighter nearby galaxy. Ground-based adaptive optics in the 2-4\,$\mu$m range and/or better signal-to-noise in the current infra-red range (using the newer Wide Field Camera 3 on {\sl HST}) should clarify the situation.

\section{Summary}
The IR-UV spectrum of \psr\ is non-monotonic, and requires a combination of emission or absorption lines and/or continuum components. The putative features might be  explained as cyclotron lines, hinting at an outer-magnetospheric electron belt. There is also the possibility of a cold ($T\sim1000$\,K) thermal component, such as  a passive, dusty circumstellar disk. Further observations in the IR are  required to confirm and probe the potential disk component.

The multi-wavelength spectrum of \psr\ suggests that a single, very broad PL component can describe the non-thermal emission surprisingly well. Interestingly, there may be evidence for  such a component  in other pulsars that have both $\gamma$-ray and IR/optical data.
A survey of  $\gamma$-ray pulsars in the IR can shed light on the ubiquity of the ultra-broad power-law magnetospheric component.

\medskip\noindent
{\bf Acknowledgements}
Based on {\sl HST} programs GO-7836, GO-9156, GO-9797, GO-10600 and
GO-11629). Support was provided by NASA through grants
from the Space Telescope Science Institute, which is operated
by the Association of Universities for Research in Astronomy,
Inc., under NASA contract NAS5-26555. We also made use of archival {\sl Chandra}
observations provided by the HEASARC service.
This work was partially supported by NASA grant
NNX09AC84G and by the Ministry of Education
and Science of the Russian Federation (contract
11.G34.310001). We are indebted to S. Zharikov
for providing the VLT spectrum. We also thank
D. Thatte of STScI for providing the dark frame
required for the NICMOS reduction.

\bibliography{database}

\appendix

\section{Data reduction details}
\subsection{STIS FUV grating}

\begin{figure*}
 \centering
\includegraphics[width=5.5in,angle=0]{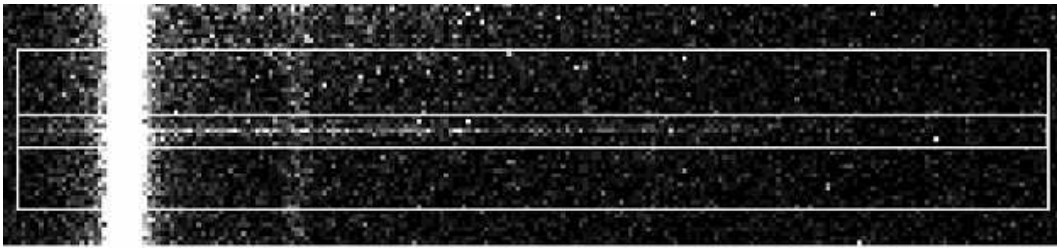}
\vspace{1.5cm}
 \caption{Raw FUV-MAMA spectrum of B0656. The boxes
show approximate regions for the source and background extraction
used in spectral analysis. } \label{raw:fuv:image}
\end{figure*}

 The source was imaged on the Far-Ultraviolet
 Multi Anode Micro-channel Array (FUV-MAMA). The
low-resolution
 grating G140L (which covers the wavelength interval $\approx1150$--1700\,\AA)
with the $52'' \times 0\farcs5$ slit was used. The data were taken
during two consecutive orbits (including the target acquisition).
The total scientific exposure time was 4950\,s \footnote{The planned
additional observations (6 orbits) were canceled because of the failure of STIS.}. FUV-MAMA was operated in TIME-TAG mode, which allows the
photon arrival times to be recorded with 125 $\mu$s
resolution\footnote{However, the absolute time of photon arrival is
not known to better than 1 s (Shibanov et al.\ 2005).}.

For each exposure, we processed the raw ``high-resolution'' images
($2048\times2048$ pixels; plate scale of $0\farcs0122$ per pixel
--- see \S11 of the STIS IHB) using the most recent calibration
files.
 As an output, we obtained flat-fielded low-resolution
($1024\times1024$ pixels; plate scale $0\farcs0244$ pixel$^{-1}$;
spectral resolution 0.58 \AA\ pixel$^{-1}$) images and used them for
the spectral analysis.

The processed images show a nonuniform detector background with a
flat (constant) component and the so-called ``thermal glow''
component \citep{1998AJ....116..789L} that dominates over most of the detector
area and grows with increasing the temperature of the FUV-MAMA
low-voltage power supply (LVPS)
(the average LVPS temperatures were 42.60 and 42.84 C in the two
consecutive orbits of our observation). The thermal
glow is the strongest in the upper-left quadrant of the detector,
where the dark count rate can exceed the nominal value,
$6\times10^{-6}$ counts s$^{-1}$ pixel$^{-1}$, by a factor of 80. To
reduce the contamination caused by the thermal glow background, the
source was imaged close to the bottom edge of the detector.

We find the B0656's spectrum centered at $Y=102\pm 2$ pixels in each
of the flat-fielded images (the centroid position slightly varies
with $X$), where $X$ and $Y$ are the image coordinates along the
dispersion and spatial axes, respectively.  Even at this location on
the detector the background still exceeds the nominal value by a
factor of 1.5--3 (typical values are
 1--2 $\times 10^{-5}$ counts s$^{-1}$ pixel$^{-1}$), depending on the position along the dispersion
axis. To improve S/N, we co-added the exposures using the STSDAS
 task ``mscombine'' (the result is shown in Fig.~\ref{raw:fuv:image}).
The $Y$-positions of the centroids differ by less than 3 pixels for
different exposures and different wavelengths ($X$-positions).

To subtract the strong, nonuniform background, we used a custom
 IDL routine with capabilities of grouping and fitting the
background and selecting an optimal extraction box size depending on
the position along the dispersion axis (see also \citealt{2004ApJ...602..327K}). The background is taken from the two strips,
$33\leq Y\leq 92$ and $113\leq Y \leq 172$, adjacent to the source
region, $93\leq Y\leq 112$. To obtain the spectrum with a
sufficiently high S/N, we have to bin the spectrum heavily; after
some experimenting, we chose 5 spectral bins ($\lambda$-bins; see
Table \ref{tab:fuv}). The bins exclude the regions contaminated by
the geocoronal emission (Ly${\alpha}$ line and the OI line at 1304 \AA).
Since the OI line at 1356 \AA\ is essentially not seen, we do not exclude
the region around 1356 \AA\ from the spectral analysis.
 The bins outside the contaminated regions were
chosen to have comparable S/N ($\approx 7$--11), whenever possible.

For each of the $\lambda$-bins, we calculate the total number of
counts, $N_t$, within the extraction boxes of different heights
(one-dimensional apertures): $A_{s}=5$, 7, 9, 7, and 7 pixels,
centered at $Y=102$ for the first two $\lambda$-bins and at $Y=103$
for the rest of the $\lambda$-bins. To evaluate the background, we
first clean the background strips (see above) from outstanding
($>10^{-3}$ cts s$^{-1}$ pixel$^{-1}$) values (``bad pixels'') by
setting them to local average values. Then, for each of the
$\lambda$-bins, we fit the $Y$-distribution of the background counts
with a first-order polynomial (interpolating across the source
region), estimate the number $N_b$ of background counts within the
source extraction aperture $A_s$, and evaluate the number of source
counts, $N_{s}=N_t-N_b$ (Table \ref{tab:fuv}).

The uncertainty $\delta N_s$ of the source counts is evaluated as
$\delta N_s = [N_s + \delta N_b^2(1+A_s/A_b)]^{1/2}$, where $\delta
N_b$ is the Poisson background uncertainty scaled to the source aperture. We binned
the distribution of background counts along the $Y$-axis with the
bin sizes equal to $A_s$ and calculated $\delta N_b$ as the
root-mean-square of the differences between the actual numbers of
background counts in the bins and those obtained from the fit to the
background.
 We calculated $\delta N_s$ and S/N for various extraction box heights
and found the $A_s$ values maximizing S/N for each $\lambda$-bin
(see Table \ref{tab:fuv}).

We calculated the average spectral fluxes in the $\lambda$-bins:
\begin{equation}
 \langle F_{\lambda}\rangle_i =
\frac{\int_{\Delta\lambda_i} R_{\lambda}\lambda\ F_{\lambda}\,{\rm
d}\lambda} {\int_{\Delta\lambda_i} R_{\lambda}\lambda\, {\rm
d}\lambda} = \frac{C_i} {\int_{\Delta\lambda_i}
 R_{\lambda}\lambda\, {\rm d}\lambda}\, ,
\end{equation}
where $C_i$ is the source count rate in the $i$-th
 $\lambda$-bin corrected for the finite size of the source extraction aperture, and
$R_{\lambda}$ is the system response that includes the Optical
Telescope Assembly throughput and accounts for the grating and slit
losses and time-dependent sensitivity losses (Bohlin, 1999; see also
\S 3.4.12 of the HST Data Handbook for STIS\footnote[6]{
http://www.stsci.edu/hst/stis/documents/handbooks/ currentDHB/STIS\_longdhbTOC.html}
for details). 

\subsection{COS spectroscopy}\label{cos_section_a}

\begin{figure*}
 \centering
\includegraphics[width=\hsize,height=0.3\hsize]{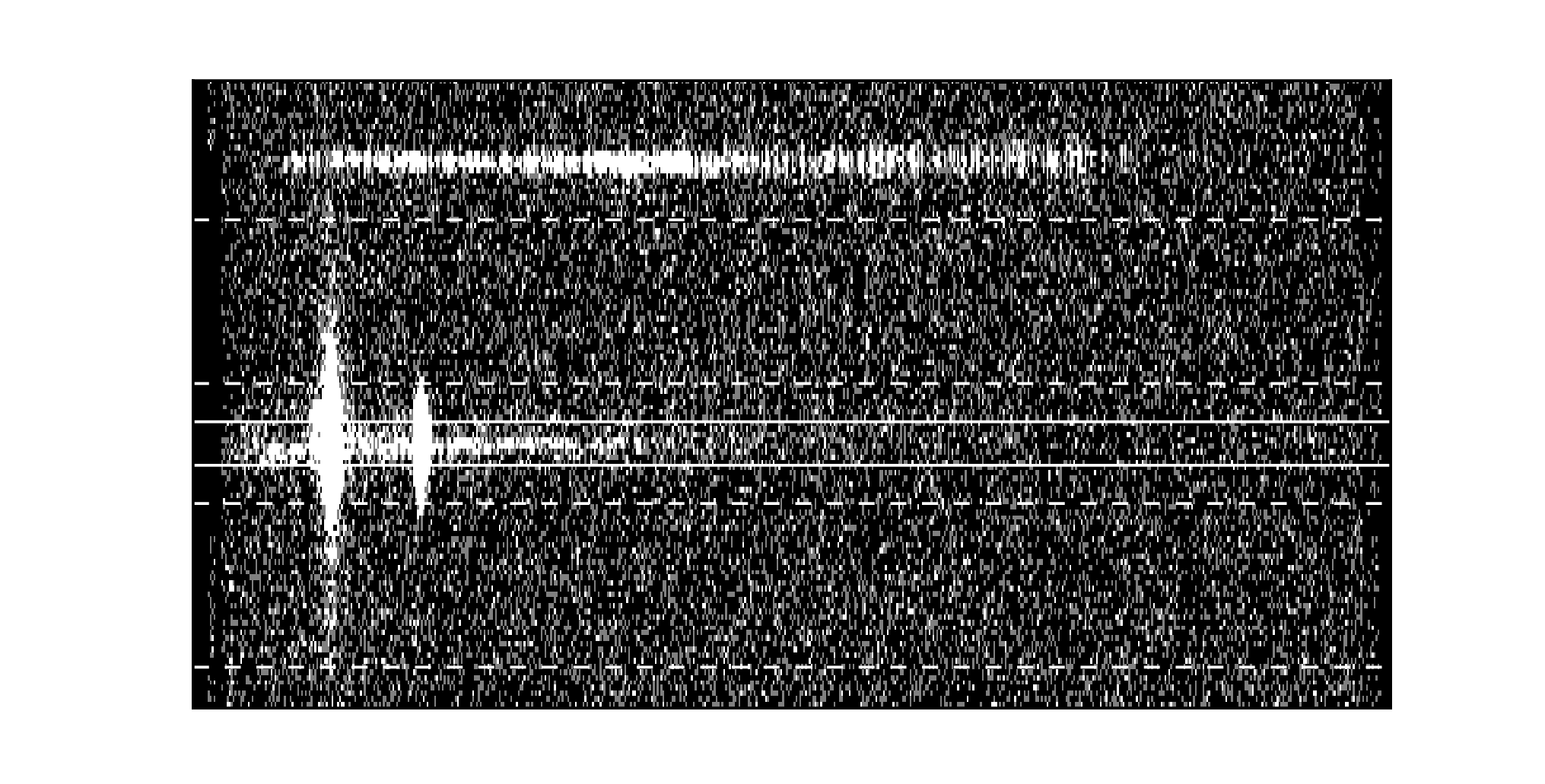}
\caption{COS image of \psr, showing events filtered by pulse height and data quality (binning for display purposes is 5$\times$2\, original pixels, 0\farcs115$\times$0\farcs184). Approximate spectral and sky extraction regions are shown (see text). The bright strip across the top of the image is the calibration spectrum, and the two prominent bright patches are images of the round aperture centered on bright geocoronal emission lines, which look elliptical after binning.}\label{raw_cos}
\end{figure*}

B0656 was observed with the Cosmic Origins Spectrograph (COS) aboard {\em HST} with the G140L grism for a total of approximately 22\,ks. With the 1105\,\AA\ central wavelength setting, the approximate sensitivity range is 1100-2250\,\AA, with resolution $R\approx5000$ and cross-dispersion scale 0.023\,\arcsec/pix. We used the standard, clear primary science aperture (PSA), a 2\farcs5 diameter circle. The FUV-A detector was operated in TIME-TAG mode, so that the arrival time (with precision of 32\,ms) and pulse-height were recorded for each photon, together with its position. The pulse height is not a measure of the photon energy, but can nevertheless be used for screening some spurious events by excluding values 1--4 and 31\footnote{\tt http://www.stsci.edu/hst/cos/documents/isrs/ ISR2010\_09(v1).pdf}. The data were processed through the standard {\tt calcos} pipeline, version 2.12,  using the latest calibration files (January 2011) to produce corrected events lists (corrtag files). The raw 2-D image of the data is shown in Figure \ref{raw_cos}.  Wavelength calibration spectra were taken during the observation, spatially offset from the source spectrum (Figure \ref{raw_cos}).

To extract fluxes from the data, we implemented an algorithm similar to the one used for STIS-FUV above. We converted corrected photon position to wavelength. Using the image in Figure \ref{raw_cos}, we found the source spectrum roughly along $y= 495$. Next, we split the data into bins of several columns each, we generated histograms of counts by y-coordinate in order to located the source centroid and measured the width by fitting a Gaussian plus constant. This provided an estimate of the position and of the extent of the spectral trace in the cross-dispersion direction as a function of wavelength, as well as a measure of the background.

In each bin we summed the total number of counts falling within an extraction region, $\pm1.7\times\sigma_\lambda$\,pix about the trace, where $\sigma_\lambda$ is the width of the  fitted Gaussian profile. The value of 1.7 was chosen by experimentation to maximize the signal-to-noise. This aperture captures 91\% of the incident light. The measured flux in the aperture is then
\begin{equation}
f_\lambda = \frac{N_\lambda - 2\times1.7\sigma_\lambda B_\lambda}{\sum_i R_{i,\lambda}}
\end{equation}
where $N_\lambda$ is the number of photons in the aperture, $B_\lambda$ is the fitted background surface density, and $R_{i,\lambda}$ is the sensitivity function at position $i$ in the dispersion direction within the bin. This flux was corrected to the standard calibration aperture (57\,pix wide), and multiplied by a factor of 1.03 to account for the slightly diminished sensitivity of COS at the time of observation compared to the benchmark of the calibration (see ISR COS 2010-15\footnote{\tt http://www.stsci.edu/hst/cos/documents/ isrs/ISR2010\_15(v1).pdf}). We assume the uncertainty in the flux is dominated by Poissonian statistics in the source and  background. Two regions of the spectrum could not be extracted due to strong geocoronal line contamination and the corresponding $\lambda$-bins were omitted.

\subsection{STIS NUV prism}

\begin{figure*}
 \centering
\includegraphics[width=5.5in,angle=0]{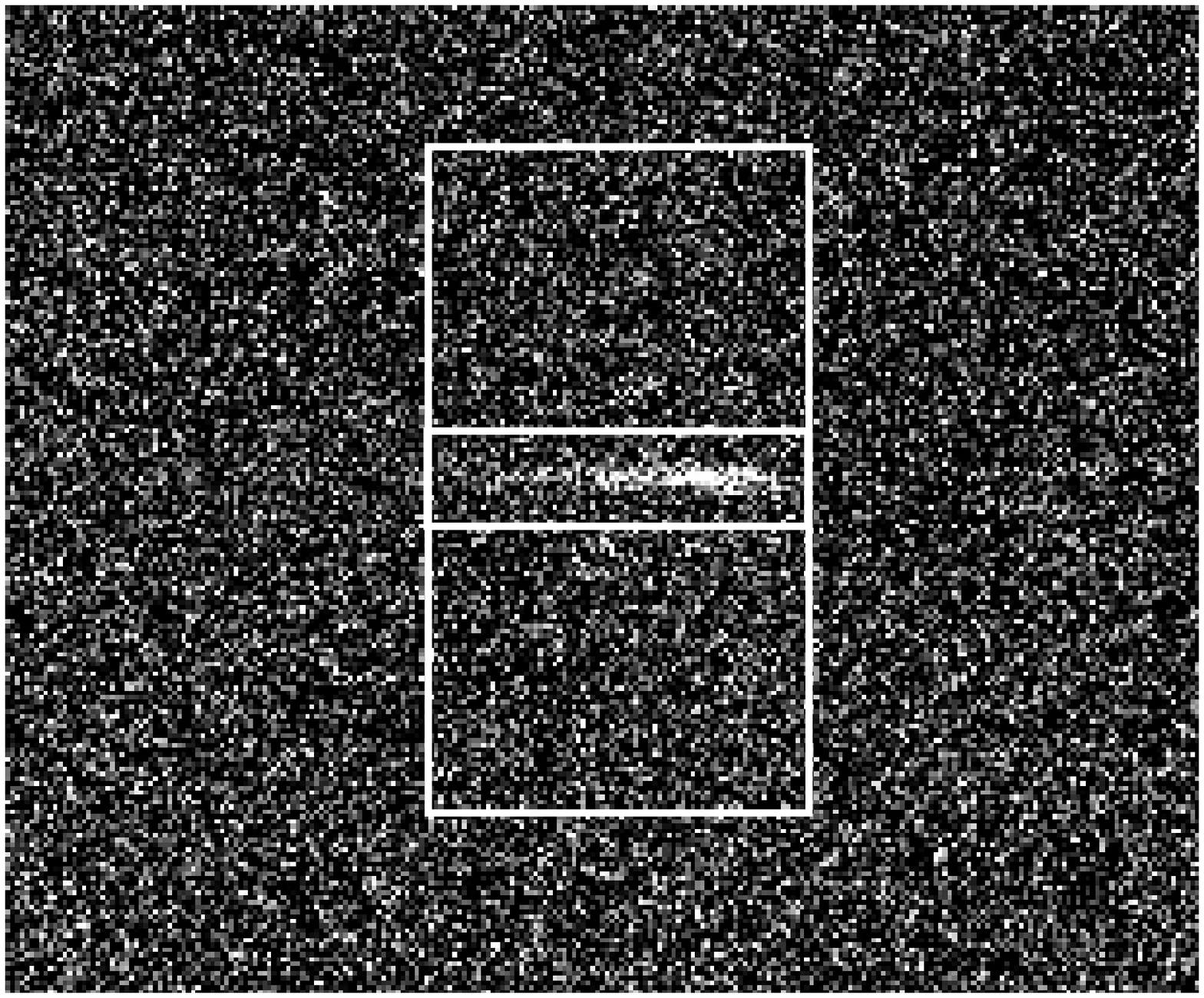}
\vspace{1.5cm}
 \caption{Raw NUV-MAMA spectrum of B0656. The boxes
show approximate regions for the source and background extraction
used in spectral analysis. }\label{raw:nuv:image}
\end{figure*}

We also retrieved the data for the program GO-9156 from the {\sl HST} archive. This program consisted of two
visits during which PSR B0656+14 was observed with STIS NUV-MAMA.
The first and second visits occurred on 2001 September 1 (start time
$52,153.94$ MJD) and  2001 November 16 (start time $52,229.47$ MJD),
respectively. The corresponding scientific exposure times were
$6790.6$ and 12761.0 seconds (3 and 5 orbits, respectively). In
these observations $52'' \times 0\farcs5$ slit together with the
PRISM (with central wavelength of 2125 \AA, pixel scale 0\farcs024) were used to obtain the
dispersed spectrum of the target.
 NUV-MAMA was also operated in
TIME-TAG mode.

For the spectral analysis we used flat-fielded low-resolution images
($1024\times1024$ pixels) and reference files provided by the STIS
pipeline.  We find the B0656's spectrum
centered at $Y=499\pm 2$ pixels in each of the flat-fielded images
(the centroid position slightly varies along the dispersion axis
$X$). To increase S/N we combined 5 exposures from the
second (longer) visit using STSDAS {\tt mscombine} task. The
resulting image is shown in Figure \ref{raw:nuv:image}. Most of the
emission is confined between $X=490$ and $X=570$.

To extract the NUV spectrum, we use the procedure similar to that
applied to the FUV data. We scan the count distribution within two
strips, $430\leq Y\leq 489$ and $510\leq Y \leq 569$, adjacent to
the source region, $490\leq Y\leq 509$. The average background count
rate at the source location is $1.2\times 10^{-3}$ counts s$^{-1}$
pixel$^{-1}$, i.e., about two orders of magnitude higher than in the
case of MAMA-FUV observation. Therefore, to obtain the spectrum with
a sufficiently high S/N,  we bin the spectrum into 8 spectral bins
with S/N$\approx 5$--11.

Contrary to grating observations, the
conversion from the pixel number (along the dispersion axis) to 
wavelength is nonlinear for
PRISM spectra\footnote{See \S12.1 of the STIS Instrument Handbook for details}. Since the
PRISM dispersion depends strongly on the wavelength (the resolution
decreases with increasing wavelength\footnote{See \S4.4 of the STIS IHB
({\tt http://www.stsci.edu/\\ hst/stis/documents/handbooks/currentIHB/\\ c04\_spectros5.html\#310764})}),
even a small shift (in terms of pixels) will translate into a large
shift in terms of the assigned wavelength at longer wavelengths
($\lambda\gtrsim3000$ \AA). Furthermore, because of the strong
dependence of the PRISM throughput on wavelength, assigning a slightly
incorrect wavelength can result in a large error in the derived flux
at $\lambda\gtrsim3000$ \AA. Hence, it is important to accurately
determine the wavelength zeropoint. The latter depends on the actual
position with respect to the aperture (slit) center as well as on
the offsets caused by the Mode Selection Mechanism (MSM)
 (positioning/tilt of the slit/PRISM) non-repeatability and thermal
drift. 

The {\sl wavecal} exposures taken between the science
exposures are expected to accurately account for the last two
effects while the source position within the slit can be measured
from the CCD images taken during each visit through the
same $52'' \times 0.5''$ slit aperture. Although at the first glance the
source appears to be well centered within the aperture,  a more detailed
investigation revealed a $0\farcs045\pm0\farcs02$ shift (about 9\%
of the slit width) of the position of the best-fit source 
centroid with respect to the slit center. This relatively small
shift amounts to $1.6$ MAMA-NUV pixels. Such a difference in
assigning the MAMA-NUV wavelength zeropoint would produce a
noticeable difference in the flux at $\lambda \gtrsim 2600$\,\AA\, because the throughput curve is steep and the resolution very low for these wavelengths. 
Hence we  account for this shift when assigning the
wavelength zeropoint. Since the measured shift has a large
associated uncertainty, we choose to ignore the spectrum at
$\lambda>2950$\,\AA\ (at these wavelengths the $0\farcs02$
uncertainty may have a large impact of the flux calibration).
Additionally, the steep throughput curve and low dispersion (compared with the line-spread function) mean that the apparent flux at wavelengths  $\lambda>$3000\,\AA\ would be contaminated by shorter wavelengths.
This effect is difficult to account
for, giving another reason to consider the flux calibration
unreliable for $\lambda>2950$\,\AA\ and limit the extraction of the NUV
spectrum to the $1790-2950$\,\AA\ range.

Aside from the caveats above, we use the same flux calibration
procedure as for the STIS FUV data to derive the NUV spectrum. 

\subsection{ACS/WFC Ramp Filter Photometry}

\begin{figure*}
\begin{center}
\includegraphics[width=0.3\hsize]{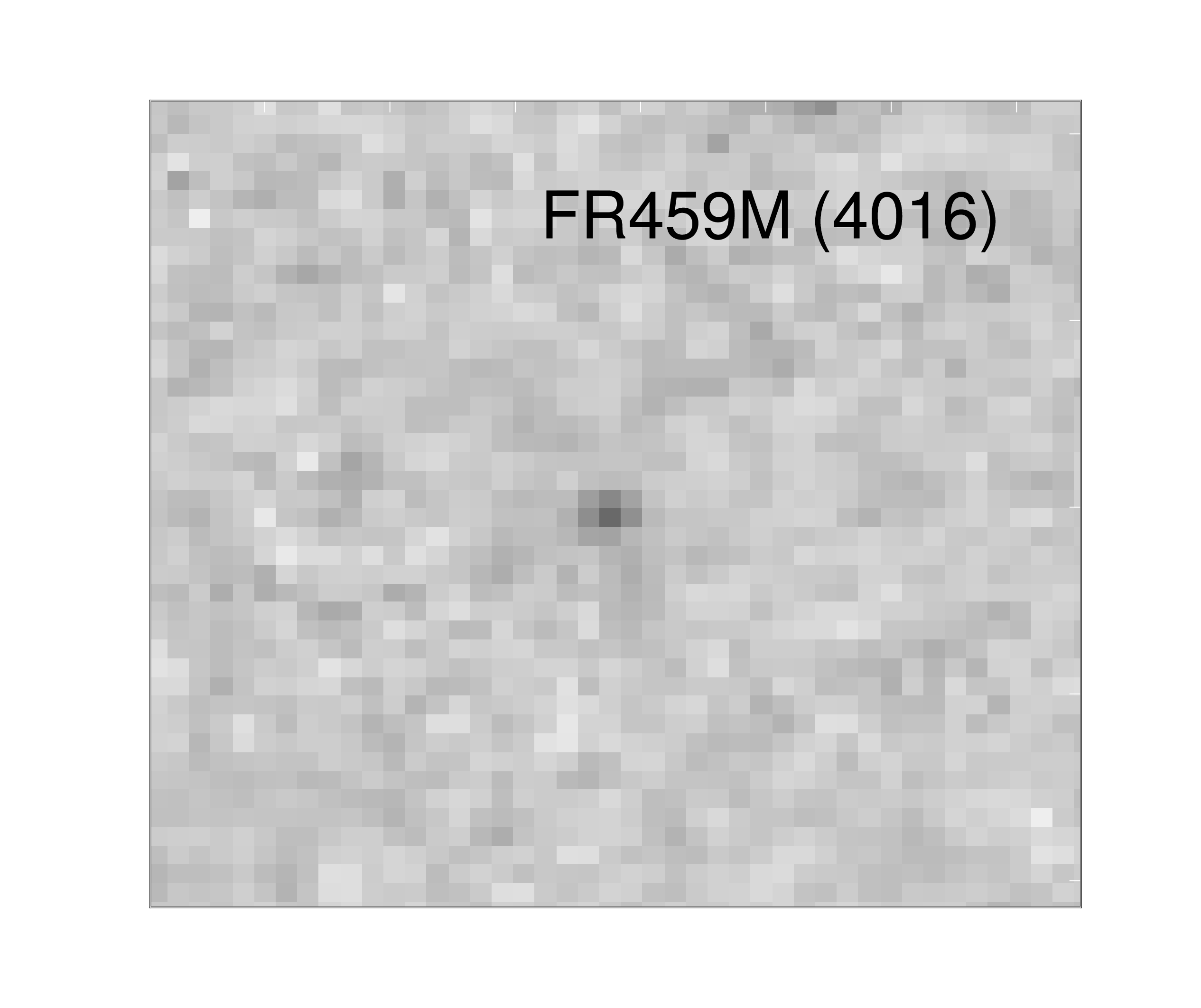}
\includegraphics[width=0.3\hsize]{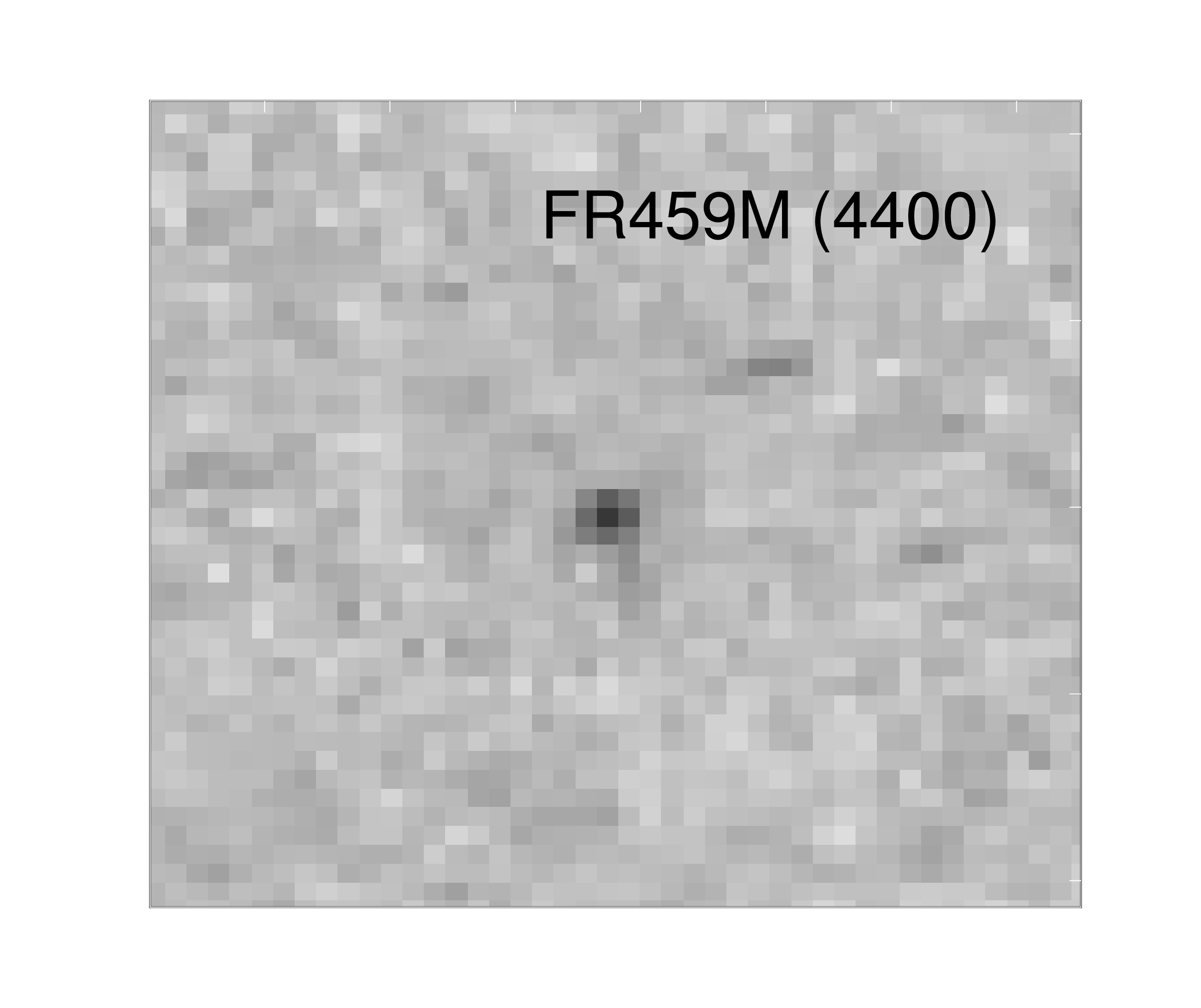}
\includegraphics[width=0.3\hsize]{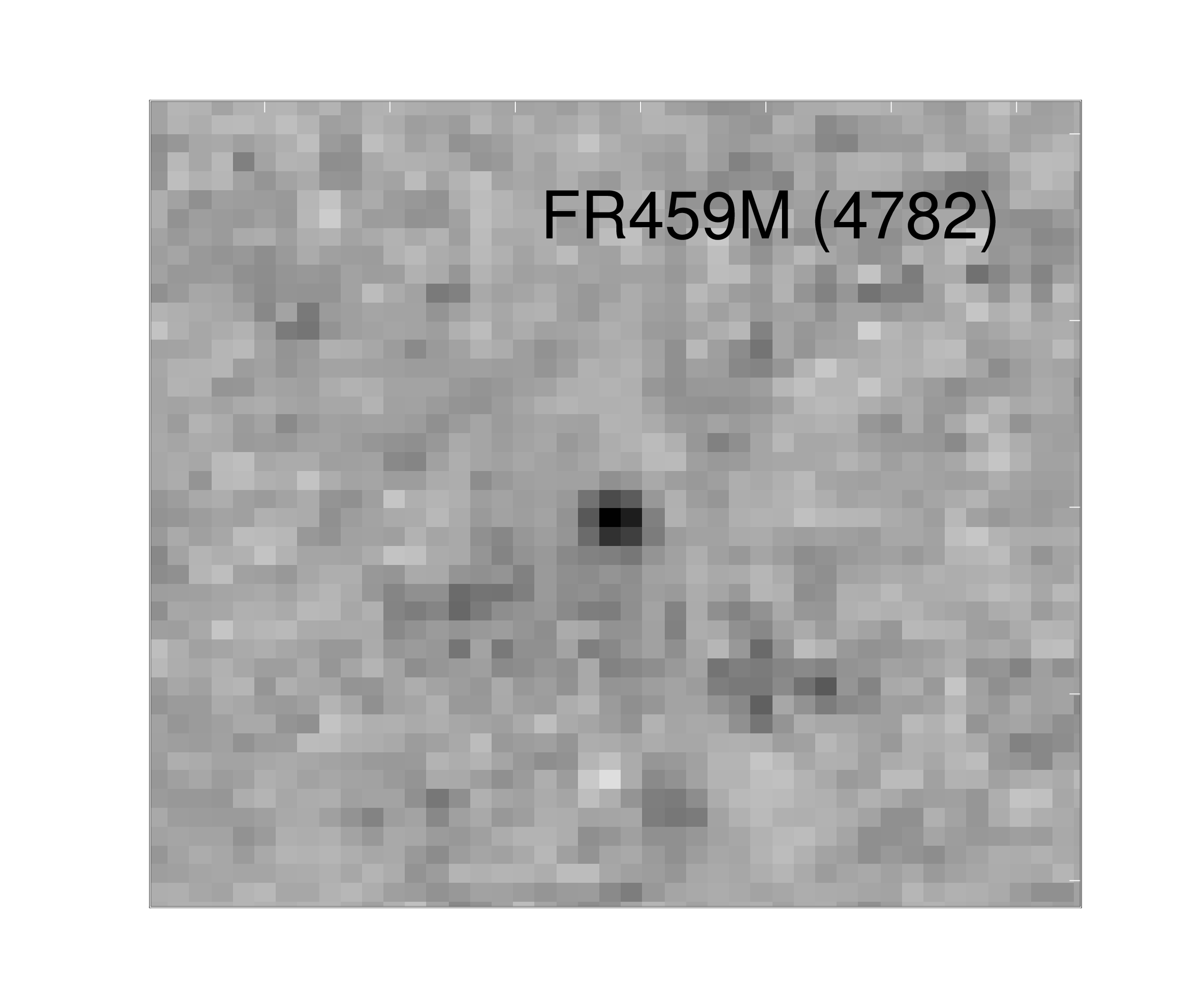}
\includegraphics[width=0.3\hsize]{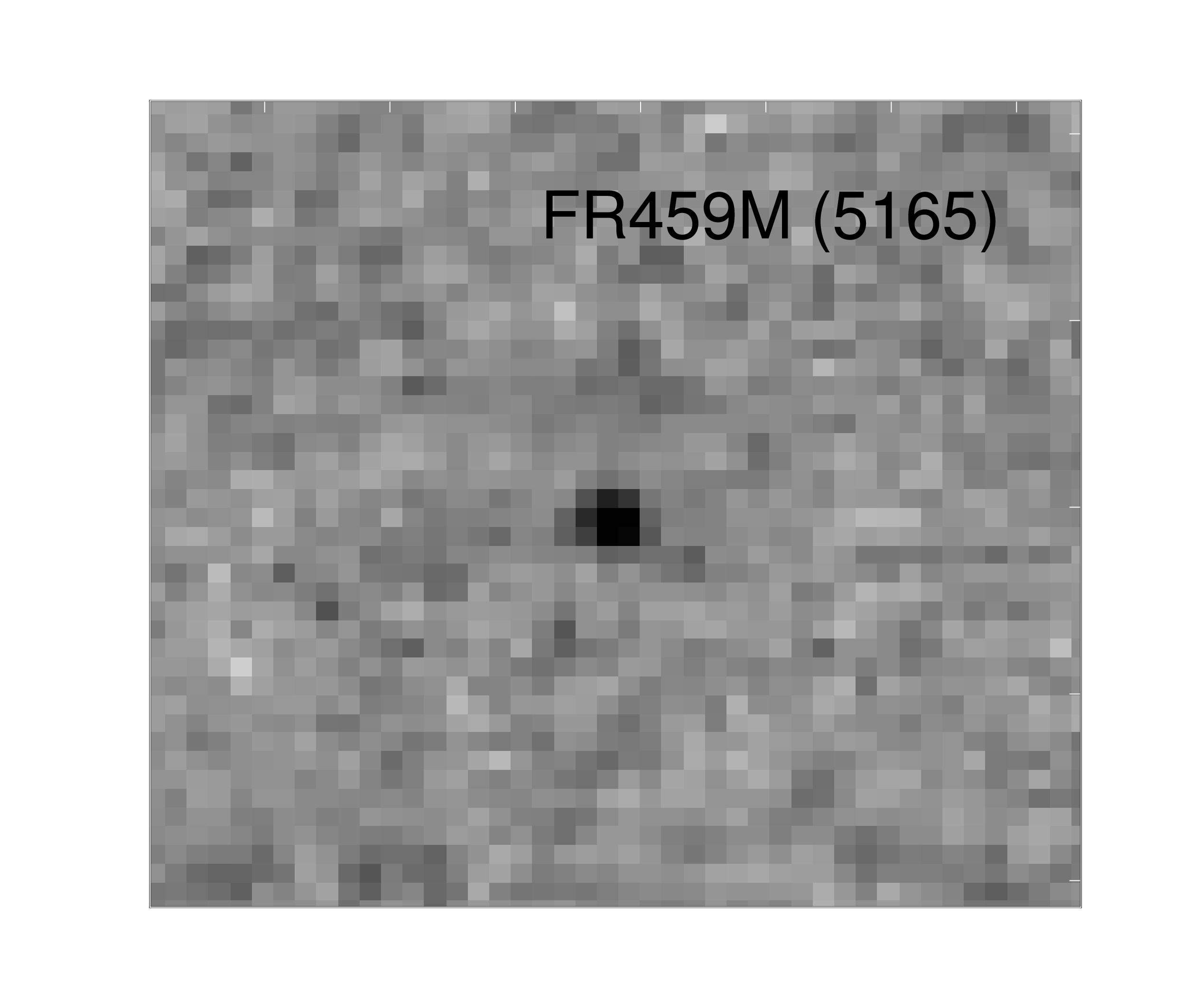}
\includegraphics[width=0.3\hsize]{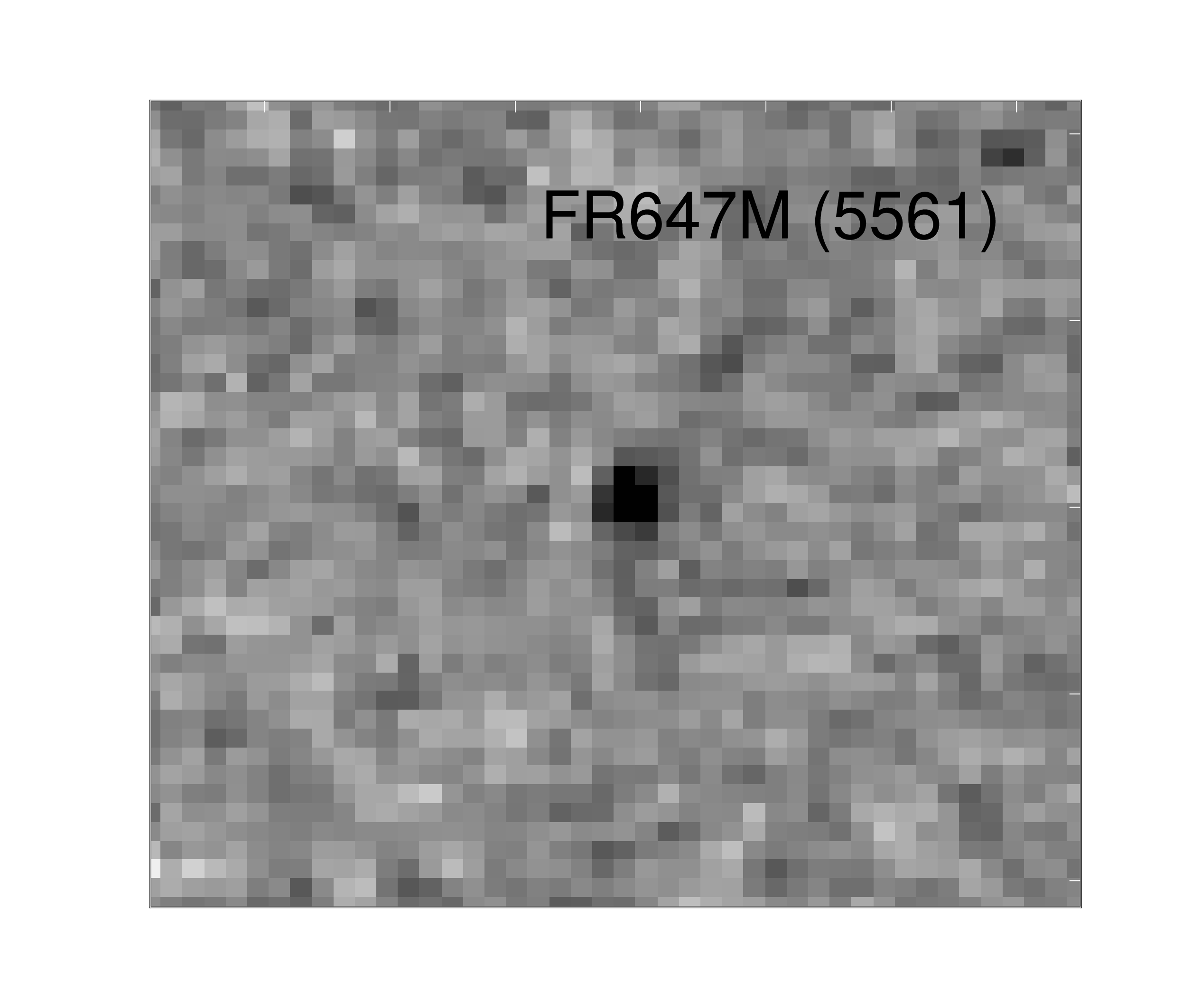}
\includegraphics[width=0.3\hsize]{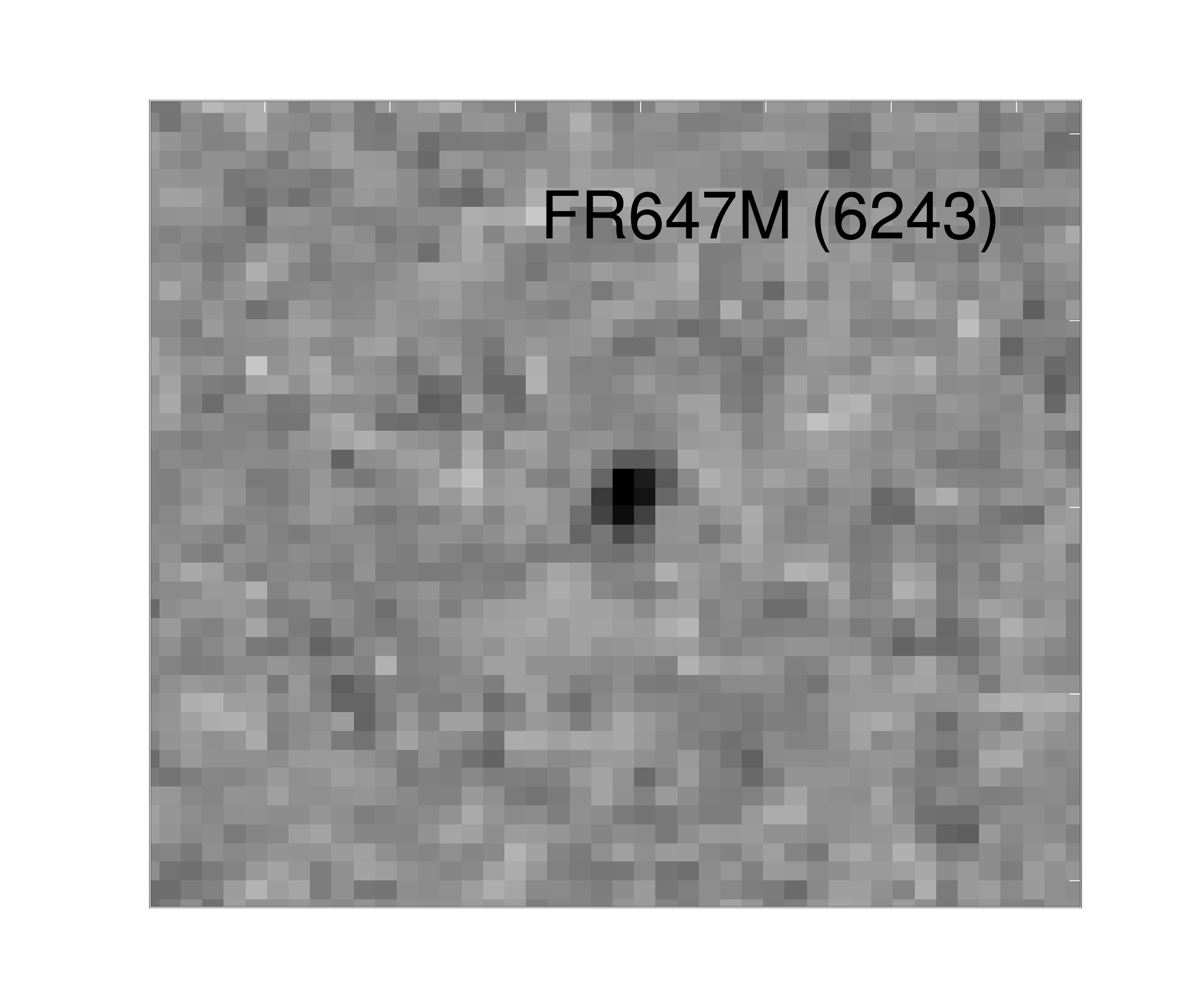}
\includegraphics[width=0.3\hsize]{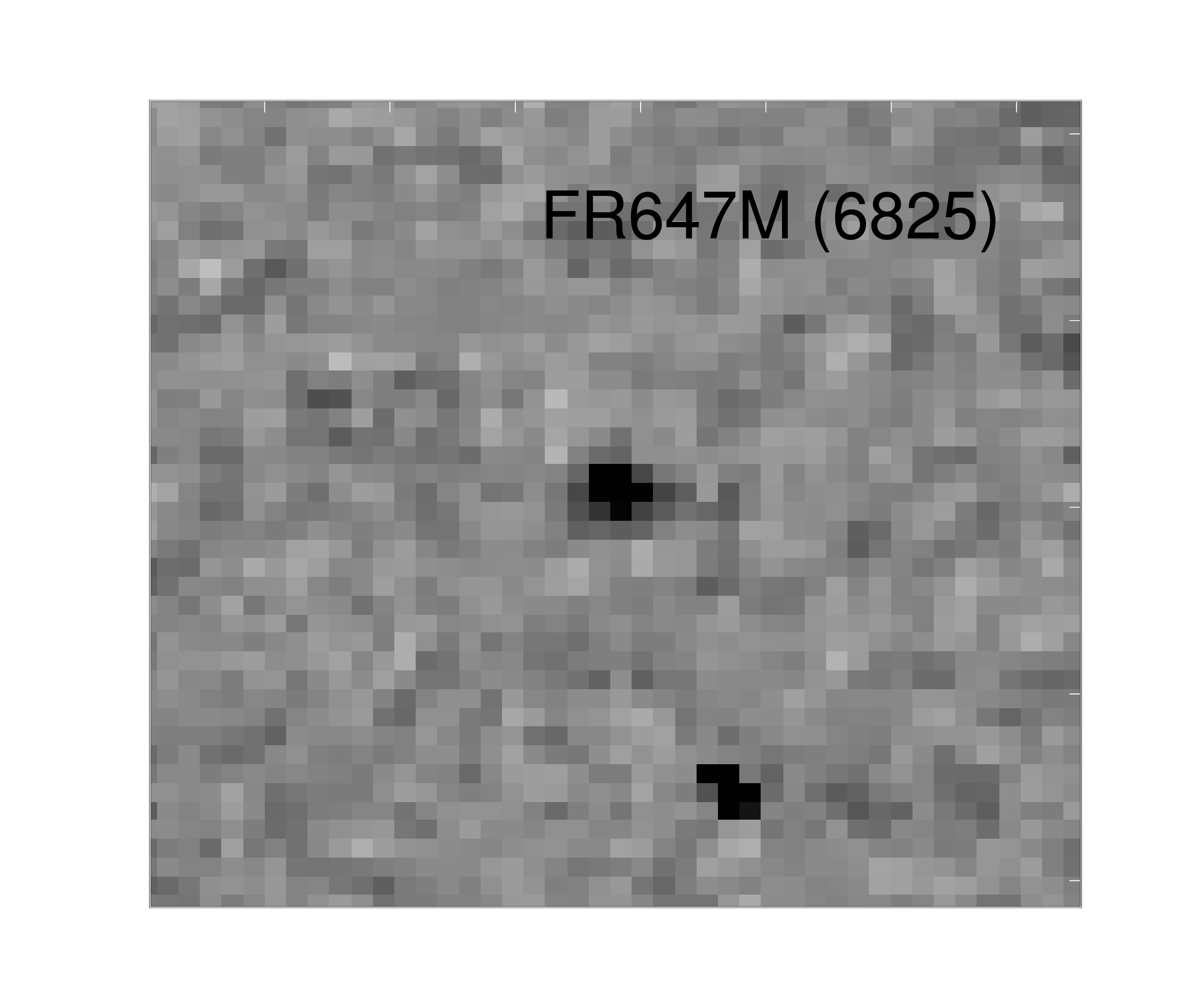}
\includegraphics[width=0.3\hsize]{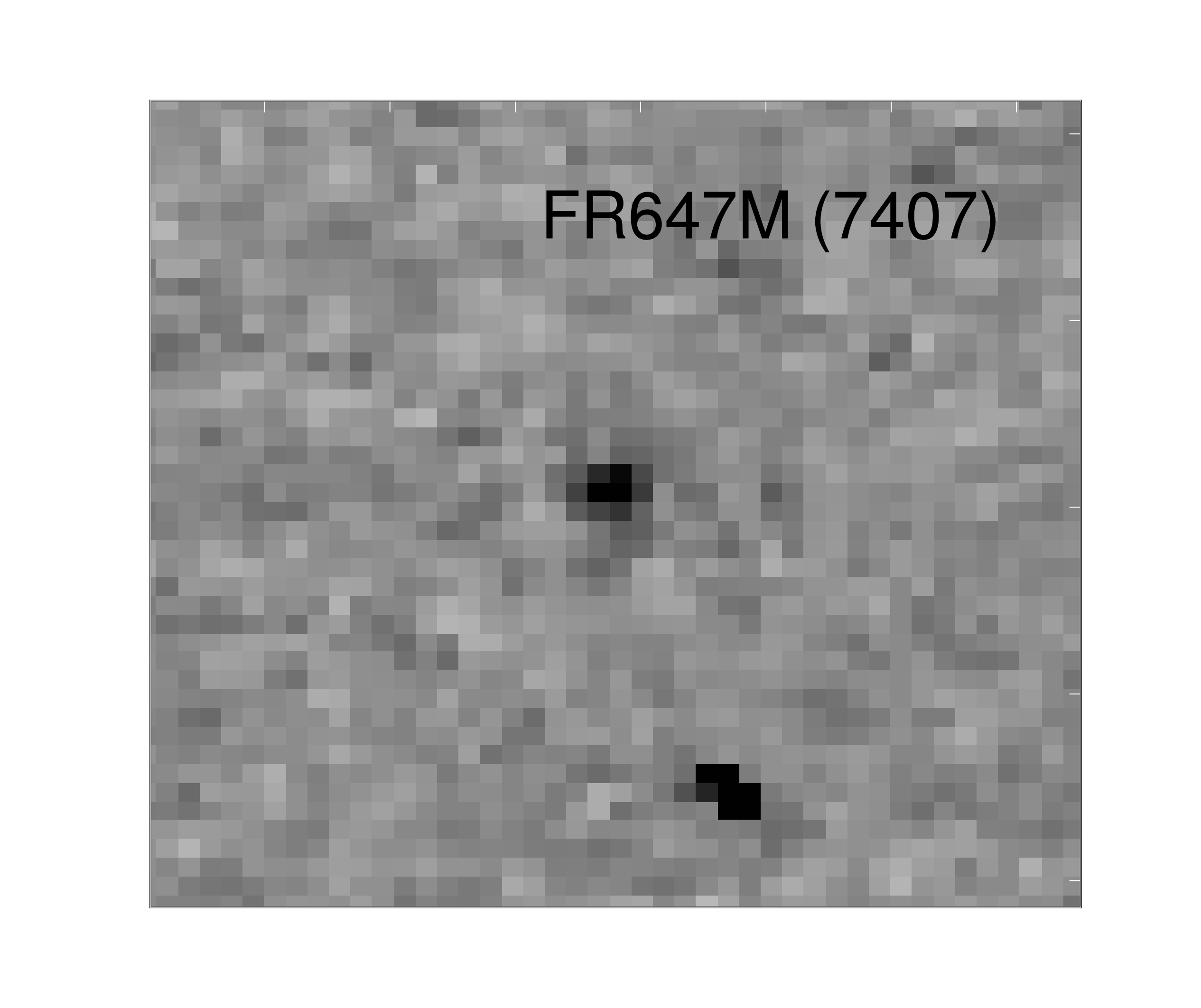}
\includegraphics[width=0.3\hsize]{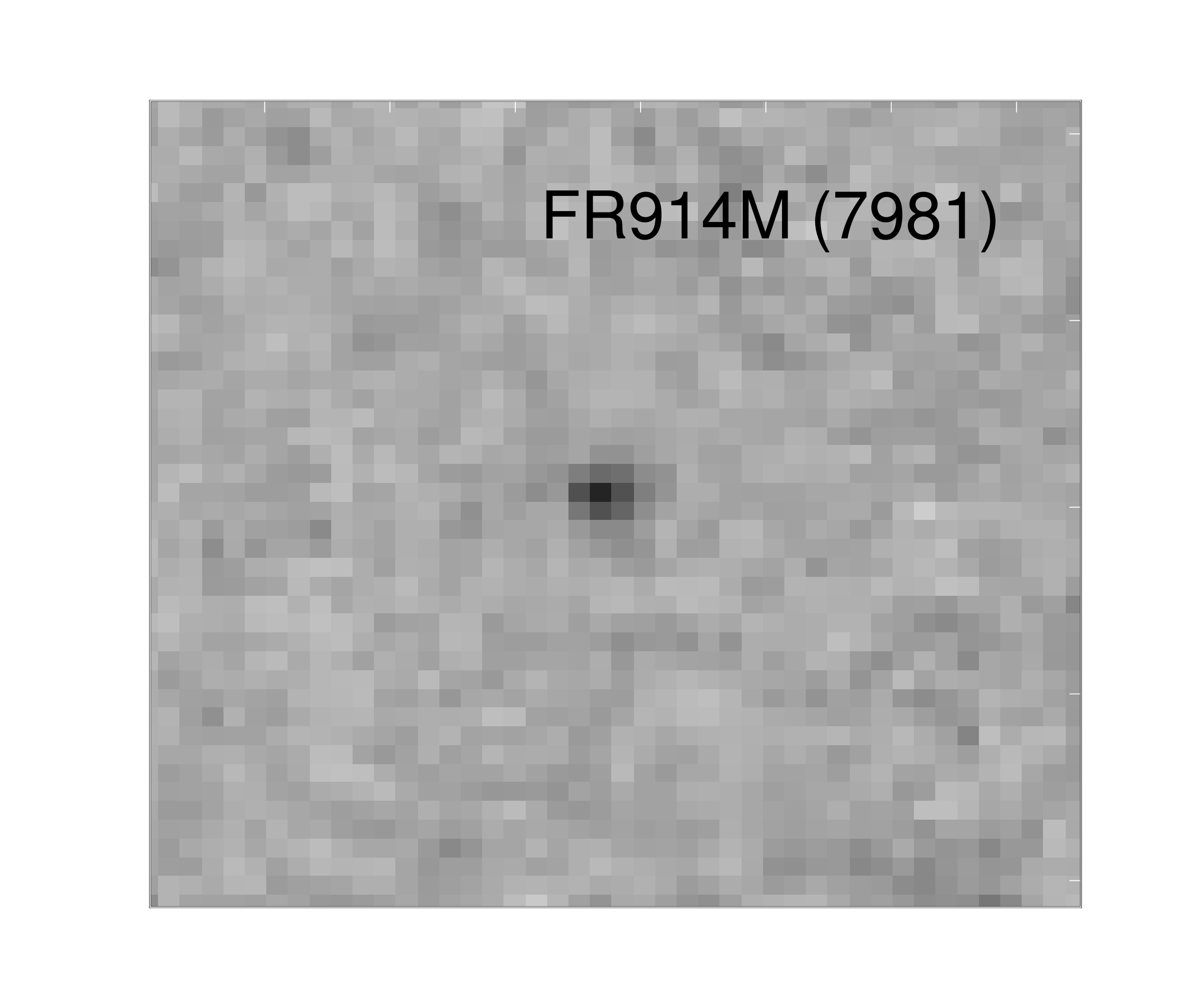}
\includegraphics[width=0.3\hsize]{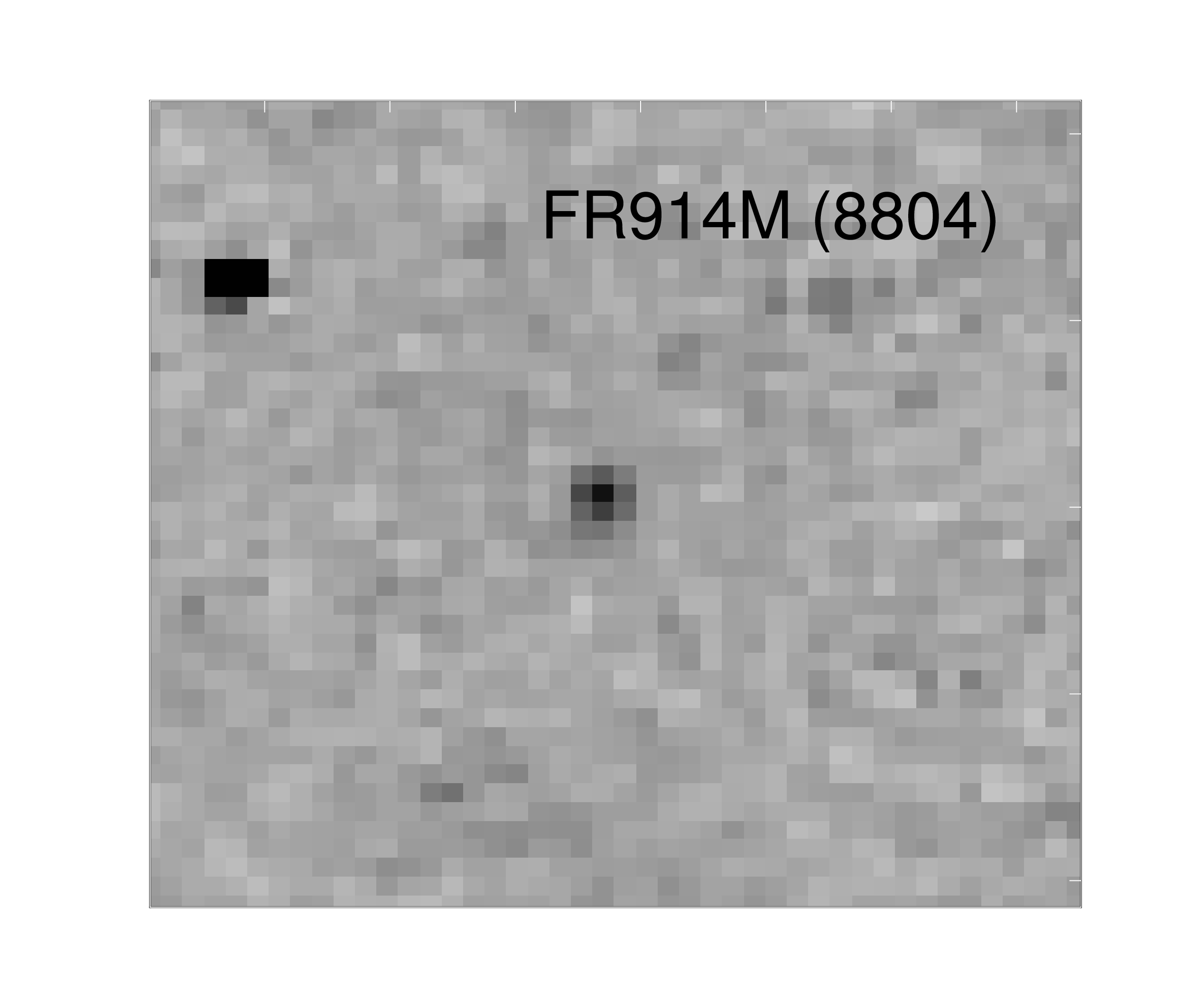}
\includegraphics[width=0.3\hsize]{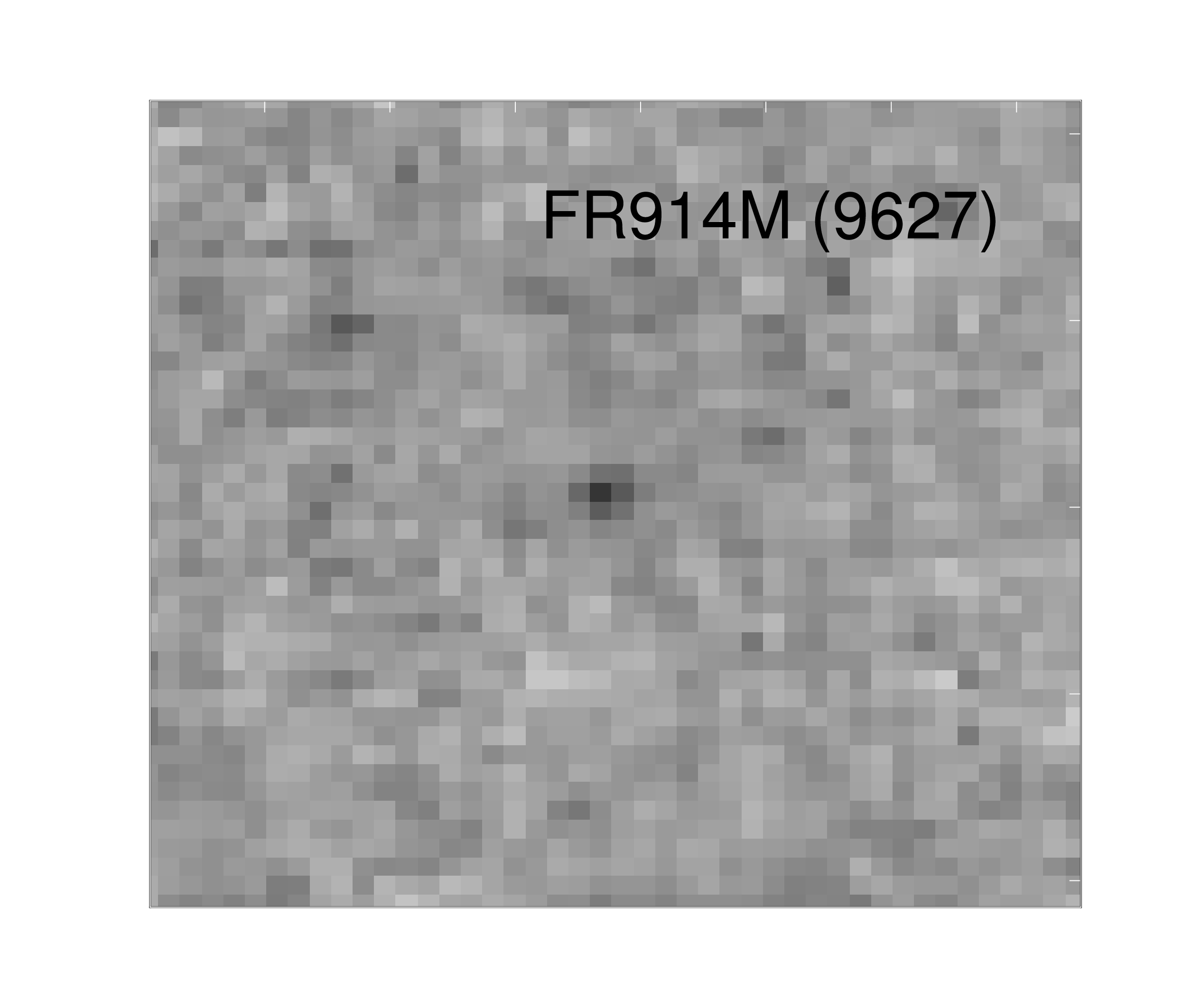}
\caption{ACS/WFC ramp filter imaging of \psr. North is up, and images are 2.2\arcsec$\times$2.2\arcsec}\label{imaging}
\end{center}
\end{figure*}

The Advanced Camera for Surveys (ACS) aboard {\sl HST} provides three ramp filters for narrow-band imaging. A ramp filter has a band-pass which depends on the part of the filter that light passes through, hence one can choose from a range of central wavelengths, with about 10\% bandwidth in each case. We utilized each of the three ramp filters, for a total of 11 photometric points, see Table \ref{log}. The pipeline-produced, drizzle-combined images are shown in Figure \ref{imaging}.

We use the same aperture size for all filters and central wavelength settings, because the width of the PSF varies only slightly with wavelength ($\sim$15\% across the optical range). We used  $r=0\farcs125$ (2.5\,pixels), which is close to the typical FWHM, and gives nearly optimal SNR in all the images. The encircled energy fraction within the aperture varies from 52\% in the red to 70\% in the blue. The background was taken from an annulus with the same center and radius 1\arcsec$<r<$1\farcs75. To estimate the aperture correction, we measured the flux within $r=0\farcs125$ and 0\farcs5 (the standard calibration aperture), for the nearby brighter star N7, located approximately 10\arcsec\ northwest of B0656 \citep{2001A&A...370.1004K}. This is one of the few field sources that has a point-source profile (most field sources are extended, i.e., galaxies), and photometry shows that it is indeed a star (see Section \ref{check}). Since the filter band-pass is position-dependent, it is necessary to choose a calibration source for  the aperture correction as close to the source of interest as possible. Finally, the fluxes were calculated using the photometric zero point delivered by the pipeline\footnote{\tt www.stsci.edu/hst/acs/documents/isrs/isr0711.pdf}. 

The CTE correction was implemented following the prescription of  \citet{2009acs..rept....1C}\footnote{\tt http://www.stsci.edu/hst/acs/documents/isrs/isr0901.pdf}.
We interpolated between the August 2005 and March 2006 values to find the parameters appropriate to our epoch of observation, giving a mean correction 
$\delta \rm{mag} = 10^a \times SKY^b \times FLUX^c \times (Y_{tran}/2000)$=0.04\,mag, where $a=0.40, b=-0.26, c=-0.47$, and the number of transfers in the $y$ direction during readout was $\rm{Y_{tran}}\approx$1100 for 4049, 4401, 4780, 5165, 7956, 8751, 9491\,\AA\ and ${\rm Y_{tran}}\approx$990 for 5566, 6244, 6822, 7393\,\AA.

\subsection{NICMOS photometry}
We downloaded the NICMOS data and the best available calibration files from the archive. Each image is a 636.1\,s  exposure in ACCUM mode. This mode is {\em non-standard}: there are multiple reads at the end of the exposure rather than reads at regular intervals throughout the exposure. This leads to inefficient screening of CRs. Four images per orbit were obtained using a dither pattern, with one orbit for F110W (2554\,s total integrations), two for F160W (5089\,s total integration) and three for F187W (7633\,s total integration - see Table \ref{log}).

\begin{figure*}
\begin{center}
\includegraphics[width=0.45\hsize]{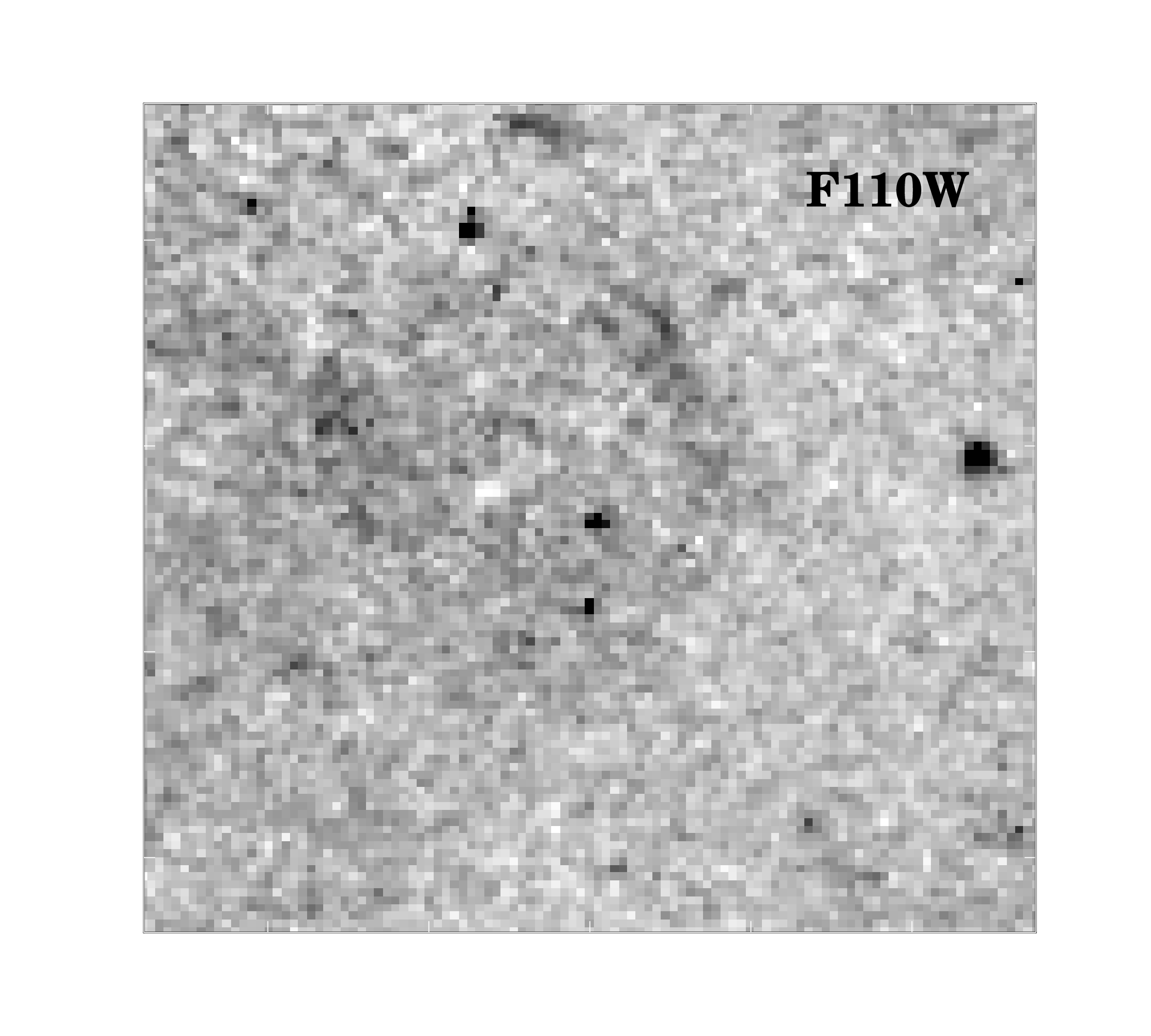}
\includegraphics[width=0.45\hsize]{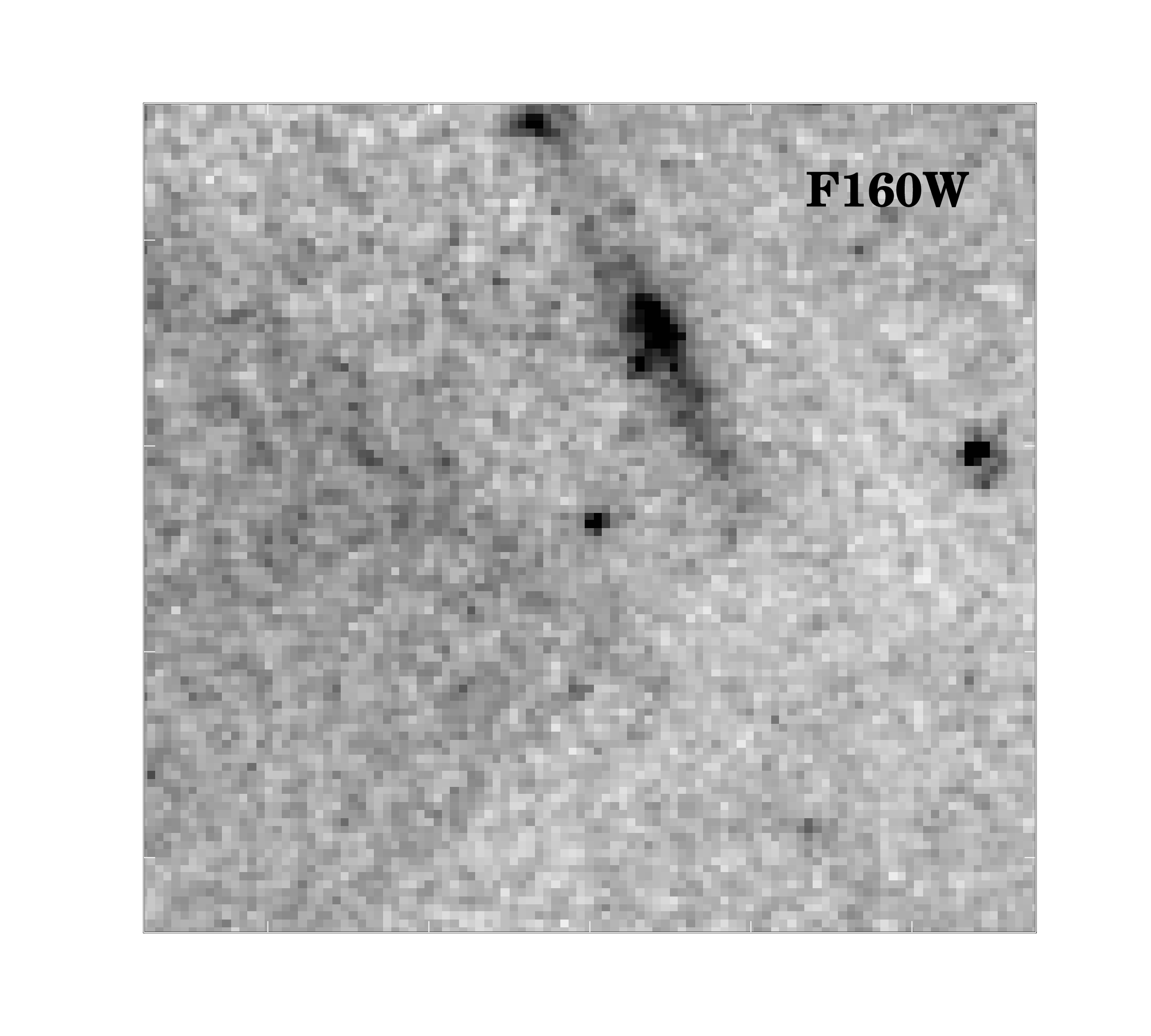}
\includegraphics[width=0.45\hsize]{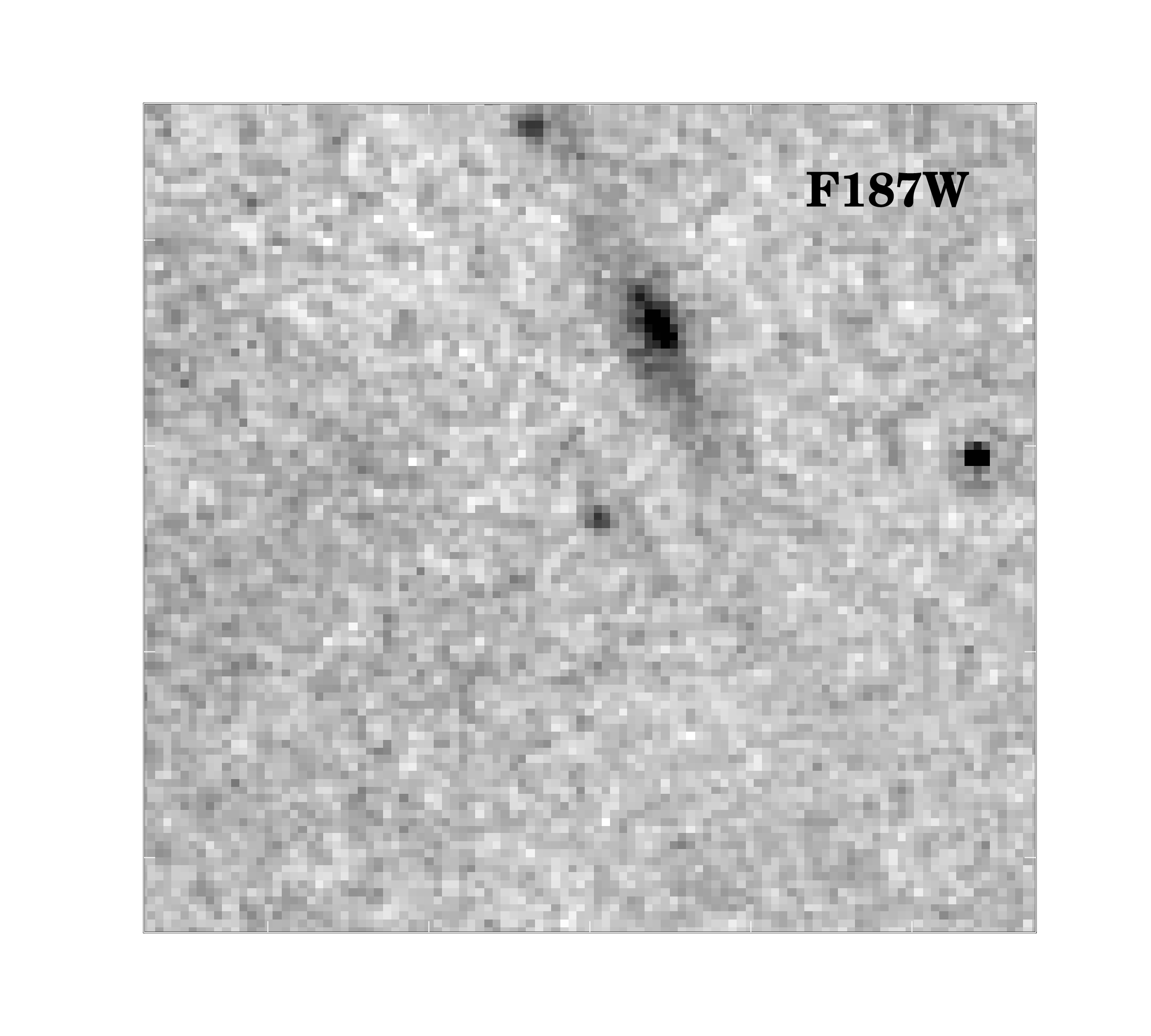}
\caption{NICMOS images of B0656. Field is centered on the pulsar, and 4\arcsec\ on each side. Compare to the images in Figure 3 of \citet{2001A&A...370.1004K}.}\label{nic_fig}
\end{center}
\end{figure*}

We processed the data, utilizing the custom-made dark file provided  by STScI at our request. This was required due to the time-variability of the detector temperature both with the age of the instrument and time ellapsed since its previous power-on. The processing of the raw data into final images shown in Figure \ref{nic_fig} included the following steps: 
\begin{itemize}
\item the pedestal correction removes bias-like additive signal from each amplifier quadrant by calculating the median pixel values, after rejection of stars and hot pixels;
\item the task {\tt multidrizzle} resamples images to a common reference frame, removing geometric distortion. It also compares the median of these images with the input resampled images, in order to identify and reject some cosmic rays/bad pixels; 
\item final images were made by a median combination. We found that the median gave better result than the average, since the ACCUM mode results in a very high incidence of cosmic rays. 
\end{itemize}

Photometry was performed using small apertures of
$r=2.5$\,pix (0\farcs19), with the aperture  correction to the 0\farcs5 nominal aperture calculated from the photometry of Star N7. The dither pattern sometimes took N7 close to the chip edge, so we used only those frames in which N7 was sufficiently far from  the edge to avoid low sensitivity pixels in the 0\farcs5 aperture. The calibrated zero point is defined relative to 1.15 times the flux in the 0\farcs5 aperture (equivalent to correcting to the ``infinite aperture''). The final flux was calculated using the header PHOTFNU keyword, provided by the pipeline.

\subsection{Optical-IR calibration check}\label{check}
\begin{figure*}
\includegraphics[width=0.8\hsize]{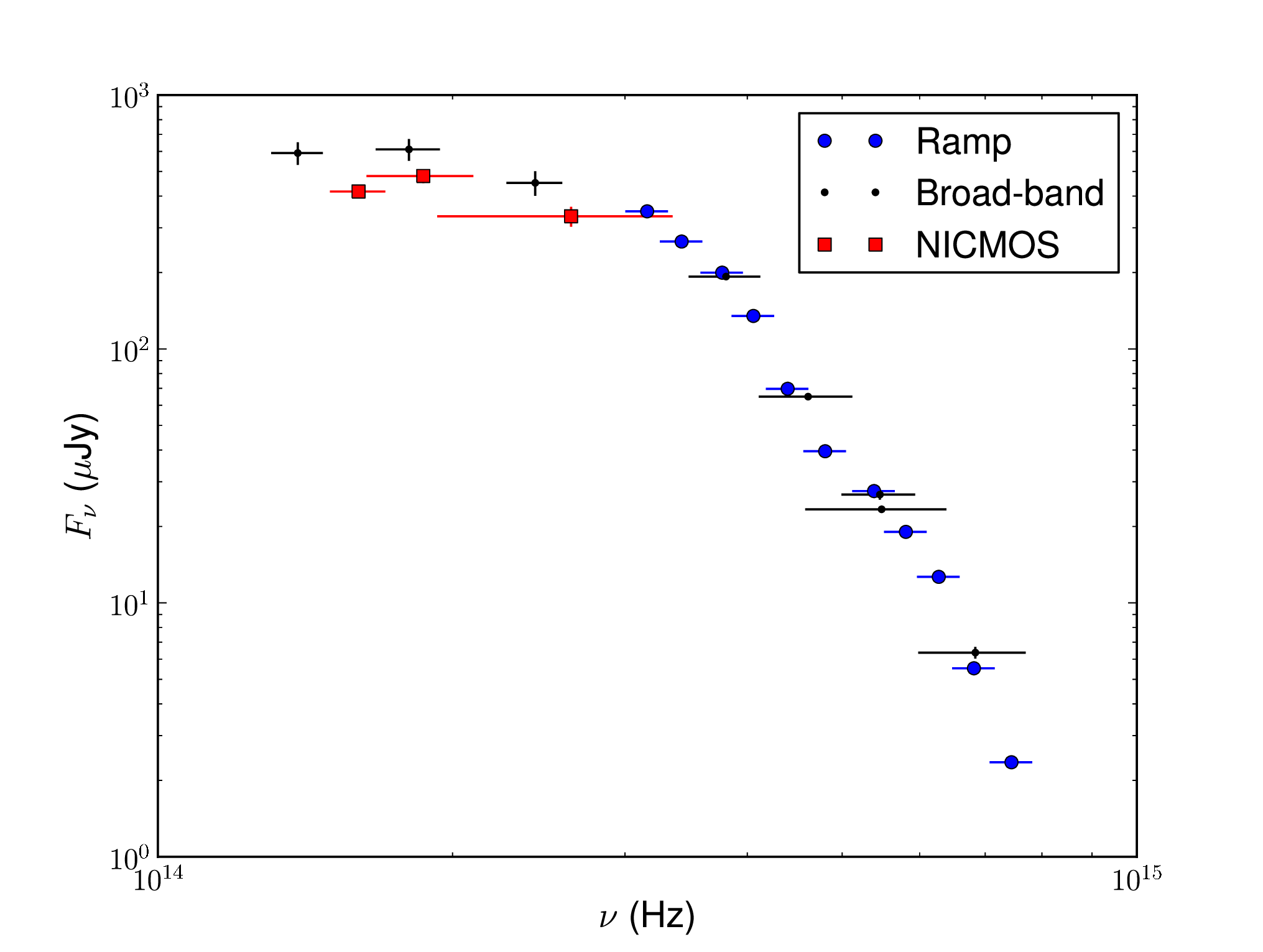}
\caption{Optical-IR spectrum of field star N7, showing 2MASS and HST broad-band photometry (black points), NICMOS photometry (red squares) and ACS narrow-band photometry (blue circles). Statistical uncertainties are quite smal in most bins, and hence not seen.}\label{star7}
\end{figure*}

In an effort to check the calibration  of our observations, we consider the flux measurements of field star N7 (see Figure \ref{star7}), also known as 2MASS J06594760+1414253, in the nominal-size aperture 0\farcs5. Using broad-band photometry ({\sl HST} in the optical and 2MASS in the NIR) to estimate the spectral class and reddening of the source\footnote{See our star type fitting tool at {\tt http://www.astro.ufl.edu/\\ $\sim$martin.durant/Sclass.html}}, we find that the star spectrum approximately matches a 4000\,K (K-type) giant at 2.5\,kpc (i.e., the Perseus Arm) reddened by $E(B-V)=0.35$ (the photometry also matches a dwarf at $\sim$80\,pc, but the reddening is unreasonably high for such a small distance). 
The broad-band fluxes and the fluxes we derive appear to match very well in the optical. 

Our IR fluxes for N7 are about 10\% higher than those of  \citet{2001A&A...370.1004K}. 
The likely  reason for the discrepancy is that that these authors also used the exposures of N7 which fell onto the low-sensitivity edges of the detector. Our IR fluxes are, however, about 20\% lower than the 2MASS fluxes. The 2MASS extraction aperture  includes the fluxes of  three faint nearby sources (resolved by {\sl HST}), which contribute roughly 10\% to the NIR flux each, and give summed fluxes consistent with the 2MASS ones, within the uncertainties.


\label{lastpage}

\end{document}